\documentclass[10pt]{article}
\usepackage{graphicx}
\usepackage{todonotes}
\usepackage{amsmath}
\usepackage{amssymb}
\usepackage{mathrsfs}
\usepackage{amsfonts}
\usepackage{bm}
\usepackage{dsfont}
\usepackage{verbatim}
\usepackage{braket}
\usepackage{hhline}   
\usepackage[absolute]{textpos}
\usepackage{xcolor}
\usepackage{appendix}
\usepackage{authblk}
\usepackage{mathrsfs}
\usepackage{nicefrac,xfrac}
\begin{document}
\title{Qudits and high-dimensional quantum computing}

\author[1]{Yuchen Wang}
\author[1]{Zixuan Hu}
\author[2]{Barry C.\ Sanders\thanks{sandersb@ucalgary.ca}}
\author[1]{Sabre Kais\thanks{kais@purdue.edu}}

\affil[1]{%
Department of Chemistry, Department of Physics and Purdue Quantum Science and Engineering Institute, Purdue University, West Lafayette, Indiana 47907, USA}
\affil[2]{%
Institute for Quantum Science and Technology,
University of Calgary, Alberta T2N~1N4, Canada}

\maketitle
\begin{abstract}

Qudit is a multi-level computational unit alternative to the conventional 2-level qubit. Compared to qubit, qudit provides a larger state space to store and process information, and thus can provide reduction of the circuit complexity, simplification of the experimental setup and enhancement of the algorithm efficiency. This review provides an overview of qudit-based quantum computing covering a variety of topics ranging from circuit building, algorithm design, to experimental methods. We first discuss the qudit gate universality and a variety of qudit gates including the pi/8 gate, the SWAP gate, and the multi-level controlled-gate. We then present the qudit version of several representative quantum algorithms including the Deutsch-Jozsa algorithm, the quantum Fourier transform, and the phase estimation algorithm. Finally we discuss various physical realizations for qudit computation such as the photonic platform, iron trap, and nuclear magnetic resonance.

\end{abstract}

\section{Introduction to qudits}
Qudit technology,
with a qudit being a quantum version of $d$-ary digits for $d>2$~\cite{brylinski2002universal}, is emerging as an alternative to qubit for quantum computation and quantum information science. Due to its multi-level nature, qudit provides a larger state space to store and process information and the ability to do multiple control operations simultaneously~\cite{Hsuan-Hao2019}. These features play an important role in the reduction of the circuit complexity, the simplification of the experimental setup and the enhancement of the algorithm efficiency~\cite{Luo2014,li2013qutrits,luo2014qudits,Hsuan-Hao2019}. The advantage of the qudit not only applies to the circuit model for quantum computers
but also applies to adiabatic quantum computing devices~\cite{ZE12,ADS13}, 
topological quantum systems~\cite{cui2015universal,cui2015universal2,Bocharov2016ternary} and more. The qudit-based quantum computing system can be implemented on various physical platforms such as photonic systems~\cite{Hsuan-Hao2019,GEZK19}, continuous spin systems~\cite{BdGS02,AHS16}, ion trap~\cite{klimov2003}, nuclear magnetic resonance~\cite{DOGRA2014,Gedik2015} and molecular magnets~\cite{leuenberger2001}. 

Although the qudit system’s advantages in various applications and potentials for future development are substantial, 
this system receives less attention than the conventional qubit-based quantum computing, and a comprehensive review of the qudit-based models and technologies is needed.
This review article provides an overview of qudit-based quantum computing covering a variety of topics ranging from circuit building~\cite{Howard2012,DWS03,Garcia-Escartin2013,Ralph2007,KNX+20},
algorithm designs~\cite{Gedik2015,Nguyen2019,NGP+20,AHS16,Cao2011,BRS17,Ivanov2012},
to experimental methods~\cite{Hsuan-Hao2019,GEZK19,BdGS02,AHS16,klimov2003,DOGRA2014,Gedik2015,leuenberger2001}. 
In this article, high-dimensional generalizations of many widely used quantum gates are presented and the universality of the qudit gates is shown.
Qudit versions of three major classes of quantum algorithms---algorithms for the oracles decision problems
(e.g., the Deutsch-Jozsa algorithm~\cite{Nguyen2019}), algorithms for the hidden non-abelian subgroup problems (e.g. the phase-estimation algorithms (PEAs)~\cite{Cao2011}) and the quantum search algorithm (e.g. Grover’s algorithm~\cite{Ivanov2012})---are discussed and the comparison of the qudit designs versus the qubit designs is analyzed.
Finally, we introduce various physical platforms that can implement qudit computation and compare their performances with their qubit counterparts. 

Our article is organized as follows.  
Definitions and properties of a qudit and related qudit gates are given in~\S\ref{sec:qudit gate}.
The generalization of the universal gate set to qudit systems and several proposed sets are provided in~\S\ref{sec:uni gate}.
Then~\S\ref{sec:gate ex}
lists various examples of qudit gates and discusses the difference and possible improvement of these gates over their qubit counterparts.
A discussion of the gate efficiency of synthesizing an arbitrary unitary~$U$ using geometric method is given in~\S\ref{sec:gategeometry}.
The next section, \S\ref{sec:qudit alg},
provides an introduction to qudit algorithms:
a single-qudit algorithm that finds the parity of a permutation in~\S\ref{sec:speedup alg},
the Deutsch-Josza algorithm in~\S\ref{subsubsec:quditDJ}, the Bernstein-Vazirani algorithm in~\S\ref{subsubsec:quditBV}, the quantum Fourier transform in~\S\ref{sec:qudit QFT}, the PEA in~\S\ref{sec:qudit PEA} and the quantum search algorithm in~\S\ref{sec:qudti search}.
\S\ref{subsec:alternative} is a section focused on the qudit quantum computing models other than the circuit model, which includes the measurement-based model in \S\ref{subsubsec:measurementbased},
the adiabiatic quantum computing in \S\ref{subsubsec:adiabatic}
and the topological quantum computing in \S\ref{subsubsec:topological}.
In~\S\ref{sec:implementationsqudits},
we provide various realizations of the qudit algorithms on physical platforms and discuss their applications. We discuss possible improvements in computational speed-up, resource saving and implementations on physical platforms. A qudit with a larger state space than a qubit can  utilize the full potential of physical systems such as photon in~\S\ref{subsec:timefreqphoton}, ion trap in~\S\ref{subsec:iontrap}, nuclear magnetic resonance in~\S\ref{subsec:nmr} and molecular magnet in~\S\ref{subsec:moluclarmagnets}. Finally, we give a summary of the qudit systems advantages and provide our perspective for the future developments and applications of the qudit in~\S\ref{SEC:qudit sum}. 

\section{Quantum gates for qudits}
\label{sec:qudit gate}
A \textit{qudit} is a quantum version of $d$-ary digits
whose state can be described by a
vector in the~$d$ dimensional Hilbert space~$\mathscr{H}_d$~\cite{brylinski2002universal}.
The space is spanned by a set of orthonormal basis vectors $\{\ket0,\ket1,\ket2,\dots \ket{d-1}\}$. 
The state of a qudit has the general form
\begin{equation}
\ket{\alpha}=\alpha_0 \ket0 +\alpha_1 \ket1 +\alpha_2 \ket2+\cdots+\alpha_{d-1} \ket{d-1}=
\begin{pmatrix}
\alpha_0 \\
\alpha_1 \\
\alpha_2 \\
\vdots   \\
\alpha_{d-1} \\
\end{pmatrix}
\in\mathbb{C}^d
\end{equation}
where $|\alpha_0|^2+|\alpha_1|^2+|\alpha_2|^2+\cdots+|\alpha_{d-1}|^2=1$.
Qudit can replace qubit as the basic computational element for quantum algorithms. The state of a qudit is transformed by qudit gates.

This section gives a review of various qudit gates and their applications.
\S\ref{sec:uni gate} provides criteria for the qudit universality and introduces several fundamental qudit gate sets.
\S\ref{sec:gate ex} presents examples of qudit gates and illustrates their advantages compared to qubit gates.  
In the last section,
\S\ref{sec:gategeometry},
a quantitative discussion of the circuit efficiency is included to give a boundary of the number of elementary gates needed for decomposing an arbitrary unitary matrix.

\subsection{Criteria for universal qudit gates}
\label{sec:uni gate}
This subsection describes the universal gates for qudit-based quantum computing and information processing. We elaborate on the criteria for universality in~\S\ref{subsubsec:universality} and give examples in~\S\ref{subsubsec:universality_eg}.

\subsubsection{Universality}
\label{subsubsec:universality}
In quantum simulation and computation, a set of matrices $U_k\in U(d^n)$ is called the universal quantum gate set if the product of its elements can be used to approximate any arbitrary unitary transformation~$U$ of the Hilbert space $\mathscr{H}^{\otimes n}_d$ with acceptable error measured in some appropriate norm~\cite{vlasov2002}. 
This idea of \textit{universality} not only applies to the qubit systems~\cite{divincenzo1995two}, but can also be extended to the qudit logic~\cite{Got99,DWS03,Zhou2003,Bullock2005,Wen-Dong2013,Mischuck2013}.
Several discussions of standards and proposals for a universal qudit gate set exist.
Vlasov shows that the combination of two noncommuting single qudit gates and a two-qudit gate are enough to simulate any unitary $U\in U(d^n)$ with arbitrary precision~\cite{vlasov2002}.
Qudit gates can themselves be reduced to, and thus simulated by, sequences of qudit gates of lower-dimensional qudit gates~\cite{Reck1994,RSdG99}
Brylinski and Brylinski prove a set of sufficient and necessary conditions for exact qudit universality which needs some random single qudit gates complemented by one two-qudit gate that has entangled qudits~\cite{brylinski2002universal}. Exact universality implies that any unitary gate and any quantum process can be simulated with zero error.
Neither of these methods is constructive and includes a method for physical implementation.
A physically workable procedure is given by Muthukrishnan and Stroud using single- and two-qudit gates to decompose an arbitrary unitary gate that operates on~$N$ qudits~\cite{Muthukrishnan2000}.
They use the spectral decomposition of unitary transformations and involve a gate library with a group of continuous parameter gates.
Brennen et al.\ identify criteria for exact quantum computation in qudit that relies on the QR decomposition of unitary transformations~\cite{Brennen2005}.
They generate a library of gates with a fixed set of single qudit operations and  "one controlled phase" gate with single parameter as the components of the universal set.
Implementing the concept of a coupling graph, they proved that by connecting the nodes (equivalently logical basis states) they can show the possibility of universal computation.

\subsubsection{Examples of universal gate sets}
\label{subsubsec:universality_eg}
An explicit and physically realizable universal set
comprising one-qudit general rotation gates and two-qudit controlled extensions of rotation gates is explained in this section~\cite{Luo2014}.
We first define
\begin{equation}
    U_d(\bm\alpha): \sum^{d-1}_{l=0}\alpha_l\ket{l}\mapsto\ket{d-1},\,
    \bm\alpha:=(\alpha_0,\alpha_1,\ldots,\alpha_{d-1}).
\end{equation}
as a~ transformation in the $d$-dimensiona that maps any given qudit state to $\ket{d-1}$.
Complex parameters of $U_d$ may not be unique and have been addressed with probabilistic quantum search algorithm~\cite{Muthukrishnan2000}.
Here in this scheme, $U_d$ can be deterministically decomposed into $d-1$ unitary transformations such that
\begin{equation}
\label{eq3.1.2}
    U_d=X_d^{(d-1)}(a_{d-1},b_{d-1})\cdots X_d^{(1)}(a_1,b_1),\;
    a_l:=\alpha_l,\, b_l:=\sqrt{\sum^{l-1}_{i=0}\alpha^2_i}
\end{equation}
with
\begin{equation}
    X_d^{(l)}(x,y)
    =\begin{pmatrix}
    \mathds1_{l-1} & & & \\
     & \frac{x}{\sqrt{|x|^2+|y|^2}}&\frac{-y}{\sqrt{|x|^2+|y|^2}} & \\
    & \frac{y^{\ast}}{\sqrt{|x|^2+|y|^2}}&\frac{x^{\ast}}{\sqrt{|x|^2+|y|^2}} & \\
    & & & \mathds1_{d-l-1}\\
\end{pmatrix}.
\end{equation}
The~$d$-dimensional phase gate is
\begin{equation}
    Z_d(\theta):=\sum^{d-1}_{l=0} \text{e}^{\text{i}(1-\text{sgn}(d-1-l))\theta}\ket{l}\bra{l},
\end{equation}
which changes  $\ket{d-1}$ by a phase $\theta$ and ignores the other states,
and $\operatorname{sgn}$ represents the sign function. 

Each primitive gate (such as $X^{(l)}_d$ or $Z_d$) has two free complex parameters to be controlled ($x,y$ in the $X_d^{(l)}$ gate and $\theta$ in the $Z_d$ gate).
Let $R_d$ represents either $X^{(l)}_d$ or $Z_d$, then the controlled-qudit
gate is
\begin{equation}
C_2[R_d]:=\begin{pmatrix}
\mathds1_{d^2-d}& \\
&R_d\\
\end{pmatrix},
\end{equation}
which is a $d^2\times d^2$ matrix that acts on two qudits. $R_d$ acts on~$d$ substates $\ket{d-1}\ket0,\ldots,\ket{d-1}\ket{d-1}$,
and the identity operation $\mathds1_{d^2-d}$ acts on the remaining substates. 

Now we work on an $N=d^n$ dimensional unitary
gate $U\in SU(d^n)$ operating on the $n$-qudit state.
The sufficiency of the gates $X^{(l)}_d, Z_d$ and $C_2[R_d]$ to construct an arbitrary unitary transformation of $SU(d^n)$ is proved in three steps. 
The first step is the eigen-decomposition of~$U$.
By the representation theory,
the unitary matrix~$U$ with $N$ eigenvalues~$\{\lambda_s\}$
and eigenstates $\ket{E_s}$ can be rewritten as
\begin{equation}
\label{eq:eig_op1}
    U =\sum^N_{j=1}\text{e}^{\text{i}\lambda_j}\ket{E_j}\bra{E_j}=\prod^N_{j=1}\Upsilon_j
\end{equation}
with eigenoperators
\begin{equation}
\label{eq:eig_op2}
    \Upsilon_j
      =\sum^N_{s=1}\text{e}^{\text{i}(1-|\operatorname{sgn}(j-s)|)
        \lambda_s}\ket{E_s}\bra{E_s}.
\end{equation}
Then the eigenoperators can be synthesized with two basic transformations as~\cite{Muthukrishnan2000}
\begin{equation}
\label{qe:eig_op3}
    \Upsilon_j
        =U^{-1}_{j,N}\,Z_{j,N}\,U_{j,N}.
\end{equation}
Here $U_{j,N}$ and $Z_{j,N}$ are the $N$-dimensional analogues of $U_d$ and
$Z_d$ such that $U_{j,N}$ is applied to the $j$th eigenstate to produce $\ket{N-1}$ and $Z_{j,N}$ modifies the phase of $\ket{N-1}$ by
the $j$th eigenphase $\lambda_j$, while ignoring all the other computation states.
According to Eq.~(\ref{eq3.1.2}),
$U_{j,N}$ can be decomposed with primitive gates $X^{(l)}_{j,N}(x,y)$. Thus, $X_{j,N}(x,y)$ and $Z_{j,N}$ are sufficient to decompose $U$.

\begin{figure}[h!]
\begin{center}
\includegraphics[width=10cm]{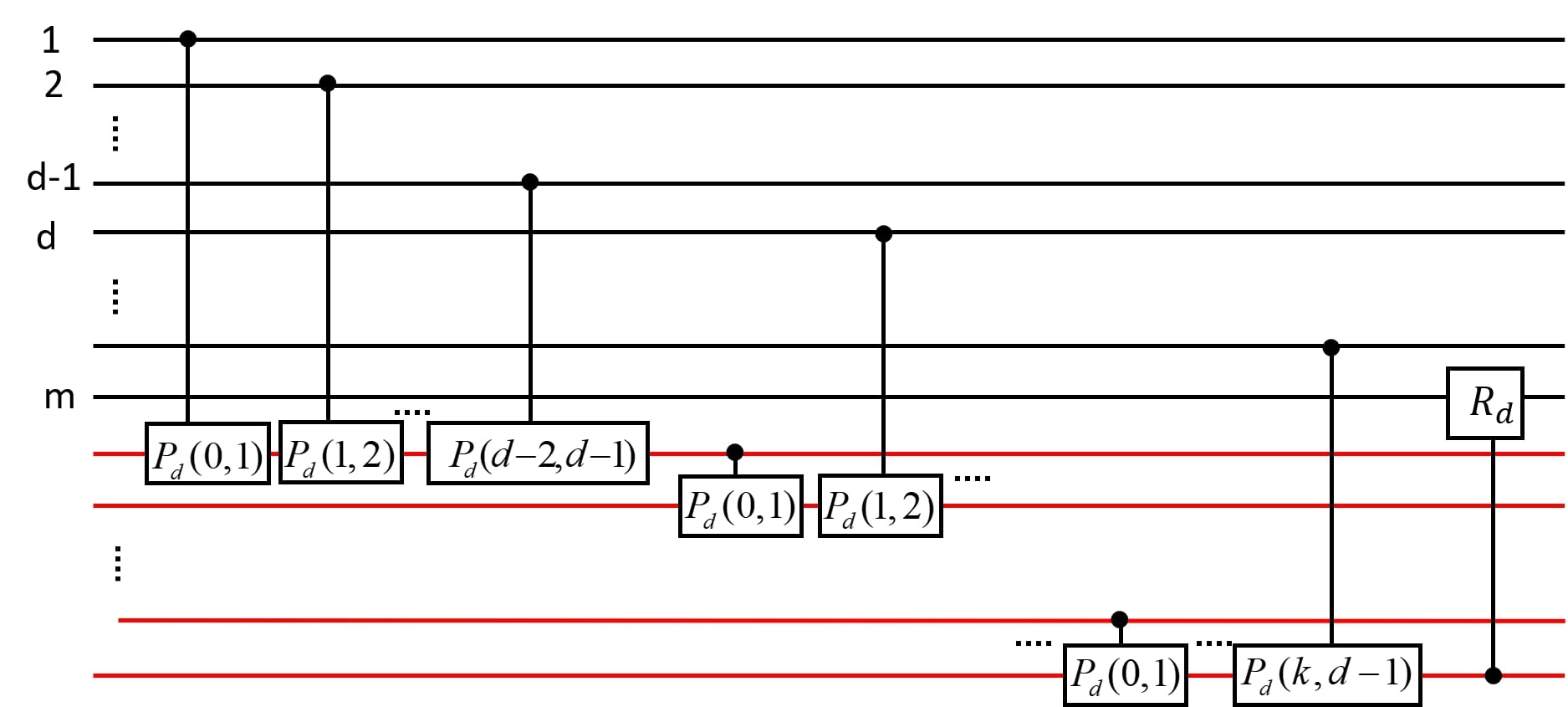}\\
\caption{The schematic circuit of $C_m[R_d]$ with $C_2[R_d]$ and $C_2[P_d(p,q)]$. The horizontal
lines represent qudits. The auxiliary qudits initialized to $\ket0$ is denoted by the red lines  and the black lines denoting $m$ controlling qudits. The two-qudit controlled gates is shown as the verticle lines. $P_d(p,q)$ is the permutation of~$\ket{p}$ and~$\ket{q}$ state, and $R_d$ is either $X^{(l)}_d$ or $Z_d$. } 
\label{fig:qudit_universal}
\end{center}
\end{figure}

The second step is decomposing
$U_{j,N}$ and $Z_{j,N}$. In other words, $U_{j,N}$ and
$Z_{j,N}$ need to be decomposed in terms of multi-qudit-controlled gates.
For convenience denote $C_m[R_d]$ as
\begin{equation}
    C_m[R_d]=\begin{pmatrix}
    \mathds1_{d^m-d}& \\
&R_d\\
\end{pmatrix},
\end{equation}
which acts on the $d^m$-dimensional computational basis of $m$-qudit
space. It is proved in the appendix of Ref.~\cite{Luo2014}  that each $U_{j,N}$ can be decomposed into some
combinations of $C_m[R_d]$ and $C_m[P_d(p,q)]$ where $P_d(p,q)$ is the permutation of~$\ket{p}$ and~$\ket{q}$ state.
The third step is using the two-qudit gates $C_2[R_d]$ and $C_2[P_d(p,q)]$ to complete the decomposition of $C_m[R_d]$.
 Fig.~\ref{fig:qudit_universal} shows a possible decomposition for $d > 2$. There are $r =\lceil (m-2)/(d-2)\rceil$ auxiliary qudits in the circuits ($\lceil x\rceil$ denotes the smallest integer greater than $x$). The last box contains $R_d=Z_d$ or~$X^{(l)}_d$. $C_m[R_d]$ is implemented with these gates combined. All of the three steps together prove that the qudit gates set
\begin{equation}
\Gamma_d:=\{X^{(l)}_d,Z_d,C_2[R_d]\}
\end{equation}
is universal for the quantum computation using qudit systems.

One advantage of the qudit model (compared to the qubit model) is a reduction of the number of qudits required to span the state space. To explain this, we need at least $n_1=\log_2 N$ qubits to represent an $N$-dimensional system in qubits while in qudits we need $n_2=\log_d N$ qudits. The qudit system has a reduction factor $k=n_1/n_2=\log_2 d$. According to Muthukrishnan and Stroud's method in Ref.~\cite{Muthukrishnan2000} a binary equivalent of their construction requires a number of qubit gates in the scale of $O(n_1^2N^2)$. By analogy, the scale of the required qudit gates using the same construction is $O(n_2^2N^2)$. So the qudit method has a $(\log_2d)^2$ scaling advantage over the qubit case.
 Furthermore, in this reviewed method, for an arbitrary unitary $U\in SU(N)$, from eq. (\ref{eq:eig_op1}) and (\ref{eq:eig_op2}) $N$ eigenoperators is needed and each can be decomposed with three rotations shown in eq. (\ref{qe:eig_op3}). Deriving from the appendix of Ref.~\cite{Luo2014}, $U_{j,N}$ can be decomposed with less than $3d^{n-1}$ multiple controlled operations. Finally, as Fig. \ref{fig:qudit_universal} has shown,$C_m[R_d]$ needs $m$ number of $C_2[R_d]$ and $C_2[P_d(p,q)]$. $U_d$ can be composed with $d-1$ numbers of $X_d^{(l)}$ as in eq.(\ref{eq3.1.2}). Therefore the total number of primitive operations $L$ in this decomposition method is
\begin{equation}
L \leqslant 2N\times 3d^{n-1}\times n\times (d-1)+N\times n\leqslant 6nd^{2n}+nd^n.
\end{equation}
It is clear that there is an extra factor of $n$ reduction in the gate requirement as the number scale of this method is  $O(nN^2)$. The other advantage is these primitive qudit gates can be easily implemented with fewer free parameters~\cite{Luo2014}.

For qudit quantum computing, depending on the implementation platform, 
other universal quantum gate sets can be considered.
For example, in a recent proposal for topological quantum computing with metaplectic anyons,
Cui and Wang prove a universal gates set for qutrit and \textit{qupit} systems,
for a qupit being a qudit with~$p$ dimensions and~$p$ is an prime number larger than $3$~\cite{cui2015universal}. The proposed universal set is a qudit analogy of the qubit universal set and it consists several generalized qudit gates from the universal qubit set.

The generalized Hadamard gate for qudits $H_d$ is
\begin{equation}
    H_d\ket{j}
        =\frac{1}{\sqrt{d}}\sum^{d-1}_{i=0}
        \omega^{ij}\ket{i},
    j\in\{0,1,2,\ldots,d-1\},
\end{equation}
where
\begin{equation}
\label{eq:omega}
    \omega:=\text{e}^{2\pi\text{i}/d}.    
\end{equation}
The  $\operatorname{SUM}_d$  gate serves as a natural generalization of the CNOT gate
\begin{equation}
   \operatorname{SUM}_d|i,j\rangle=|i,i+j(\operatorname{mod}d)\rangle,i,j\in\{0,1,2,\ldots,d-1\}.
\end{equation}
 The Pauli $\sigma_z$, with the $\pi/8$ gate as its 4th root, can be generalized to $Q[i]$ gates for qudits,
\begin{equation}
    Q[i]_d\ket{j}=\omega^{\delta_{ij}}\ket{j},
\end{equation}
with~$\omega$ defined by Eq.~(\ref{eq:omega})
and the related $P[i]$ gates are
\begin{equation}
    P[i]_d\ket{j}=(-\omega^2)^{\delta_{ij}}\ket{j}, \;\, i,j\in\{0,1,2,\ldots,d-1\}.
\end{equation}
In general $Q[i]_p$ is always a power of $P[i]_p$ if $p$ is an odd prime.

The proposed gate set for the qutrit system is the sum gate $\operatorname{SUM}_3$, the Hadamard gate $H_3$ and any gate from the set $\{P[0]_3,P[1]_3,P[2]_3\}$. As an analogue of the standard universal set for qubit $\{\operatorname{CNOT},H,T=\pi/8\mathrm{-gate}\}$, the qutrit set generate the qutrit Clifford
group whereas the qubit set generate the qubit Clifford group (the definition of the Clifford group can be found in~\S\ref{subsubsec:quditpi8}).
Whereas the rigorous proof can be found in Ref.~\cite{cui2015universal}, the proving process follows the idea introduced in Ref.~\cite{brylinski2002universal} that the gate $\operatorname{SUM}_3$ is imprimitive,
and the Hadamard $H_3$ and any gate from $\{P[0]_3,P[1]_3,P[2]_3\}$
generates a dense subgroup of $SU(3)$. Similarly, the proposed gate set for the
qupit system is  the sum gate $\operatorname{SUM}_p$, the Hadamard gate $H_p$ and the gates
$Q[i]_p$ for $i\in[p-1]$. The proof is analogous to that of the qutrit set. The Hadamard $H_p$ and the $Q[i]$ gates are combined to form a dense subgroup of $SU(p)$ and $\operatorname{SUM}_p$ is shown to be imprimitive.
Implementing Theorem 1.3 in Ref.~\cite{brylinski2002universal}, the set is a universal gate set.  These universal gate sets for the qudit systems, with fewer numbers of gates in each set compare to that in the previous examples, have the potential to perform qudit quantum algorithms on the topological quantum computer.

\subsection{Examples of qudit gates}
\label{sec:gate ex}
In this section we introduce the qudit versions of many important quantum gates and discuss some of the gates' advantages compared to their qubit counterparts.
The gates discussed are the qudit versions of the~$\pi/8$ gate
in~\S\ref{subsubsec:quditpi8}, 
the SWAP gate in~\S\ref{subsubsec:swap} and the multi-level controlled gate in~\S\ref{sec:qudit_MVCG}.
in~\S\ref{sec:improv_toffoli_gate}, we also introduce how to simplify the qubit Toffoli gate by replacing one of the qubit to qudit. This gives ideas about improving the qubit circuits and gates by introducing qudits to the system.
\subsubsection{Qudit versions~$\pi/8$ gate}
\label{subsubsec:quditpi8}
The qubit~$\pi/8$ gate $T$ has an important role in quantum computing and information processing. This gate has a wide range of applications because it is closely related to the Clifford group but does not belong to the group. From the Gottesman-Knill theorem~\cite{Got98} it is shown that the Clifford gates and Pauli measurements only
do not guarantee universal quantum computation(UQC). The~$\pi/8$ gate, which is non-Clifford and from the third level of the Clifford hierarchy,
is the essential gate to obtaining UQC~\cite{BOYKIN2000}.
This gate can be generalized to a~$d$ dimensional qudit system, where, throughout the process,
$d$ is assumed to be a prime number greater than $2$~\cite{Howard2012}. 
 
To define the Clifford group for a $d$-dimensional qudit space,
we first define the Pauli $Z$ gate and Pauli $X$ gate.
The Pauli $Z$ gate and Pauli $X$ gate  are generalized to~$d$ dimension in the matrix forms~\cite{PZ88,GKP01,BdGS02,NBD+02}
\begin{equation}
\label{eq:dXdZ}
X_d=\begin{pmatrix}
    0 & 0 &  \cdots &0 & 1  \\
    1 & 0 &  \cdots &0 & 0  \\
    0 & 1 &  \cdots &0 & 0  \\
    \vdots & \vdots & \ddots & \vdots & \vdots  \\
    0 & 0 & \cdots & 1 & 0  
\end{pmatrix},\,
Z_d=\begin{pmatrix}
    1 & 0 & 0 & \cdots & 0  \\
    0 & \omega & 0 & \cdots & 0  \\
    0 & 0 & \omega^2 & \cdots & 0  \\
    \vdots & \vdots & \vdots & \ddots & 0  \\
    0 & 0 & 0 & \cdots & \omega^{d-1}  
\end{pmatrix}
\end{equation}
for~$\omega$ the $d^\text{th}$ root of unity~(\ref{eq:omega}).
The function of the $Z$ gate is adding different phase factors to each basis states and that of the $X$ gate is shifting the basis state to the next following state. Using basis states the two gates are
\begin{equation}
\label{eq:ZjXj}
    Z_d\ket{j}:=\omega^j\ket{j}\; X_d\ket{j}:=\ket{j+1},\;
    j\in\{0,1,2,\ldots,d-1\}.
\end{equation}
In general,  we define the displacement operators as products of the Pauli operators,
\begin{equation}
\label{eq:displacement}
D_{(x|z)}=\tau^{xz}\,X_d^x\,Z_d^z, \; \tau:=\text{e}^{(d+1)\pi\text{i}/d},
\end{equation}
where $(x|z)$ correspond to the $x$ and $y$ in the exponent of $\tau$, $X$ and $Z$.
This leads to the definition of the Weyl-Heisenberg group (or the generalized Pauli group) for a single qudit as~\cite{PZ88,GKP01,BdGS02,NBD+02}
\begin{equation}
\label{eq:generalizedPauligroup}
    \mathcal{G}=\{\tau^c D_{\vec{\chi}}|\vec{\chi}\in\mathbb{Z}^2_d,c\in \mathbb{Z}_d\}\;(\mathbb{Z}_d=\{0,1,\ldots,d-1\}),
\end{equation}
where $\vec{\chi}$ is a two-vector with elements from $\mathbb{Z}_d$. With these preliminary concepts defined in Eqs.~(\ref{eq:dXdZ}) 
through~(\ref{eq:generalizedPauligroup}),
we now define the Clifford group as the following: the set of the operators that maps the Weyl-Heisenberg group onto itself under conjugation is called the \textit{Clifford group}~\cite{NBD+02,Web16},
\begin{equation}
\mathcal{C}=\{C\in U(d)|C\mathcal{G}C^\dagger=\mathcal{G}\}.
\end{equation}
A recursively defined set of gates,
the so-called Clifford hierarchy,
was introduced by Gottesman and Chuang
as
\begin{equation}
\mathcal{C}_{k+1}=\{U|U\mathcal{C}_1U^\dagger\subseteq\mathcal{C}_k\},
\end{equation}
for~$\mathcal{C}_1$ the Pauli group~\cite{Gottesman1999}.
The sets $\mathcal{C}_{k\geq3}$ do not form groups,
although the diagonal subsets of $\mathcal{C}_3$, which is our focus here,
do form a group.

The following derivations follow those in Ref.~\cite{Howard2012}. The explicit formula for building a Clifford unitary gate with 
\begin{equation}
    F=\begin{pmatrix}
\alpha&\beta\\
\gamma&\delta
\end{pmatrix}\in \mathrm{SL}(2,\mathbb{Z}_d),\;
\vec{\chi}=\begin{pmatrix}
x\\
z
\end{pmatrix}\in \mathbb{Z}_d^2
\end{equation}
is
\begin{equation}
\label{eq:CF}
    C_{(F|\vec{\chi})}=D_{(x|z)}V_F,
\end{equation}
\begin{equation}
\label{eq3.2.2}
 V_F =
  \begin{cases}
    \frac{1}{\sqrt{d}}\sum^{d-1}_{j,k=0}\tau^{\beta^{-1}(\alpha k^2-2jk+\delta j ^2)}\ket{j}\bra{k},       & \beta\neq 0\\
    \sum^{d-1}_{k=0}\tau^{\alpha\gamma k^2}|\alpha k\rangle\bra{k},  & \beta=0.
  \end{cases}
\end{equation}
The special case $\beta=0$ is particularly relevant to the later derivation,
and
\begin{equation}
\label{eq3.2.6}
    \begin{split}
 \det\left( \sum^{d-1}_{k=0}\tau^{\alpha\gamma k^2}| k\rangle\bra{k}\right) =& \tau^{\frac{\alpha\gamma}{6}(2d-1)(d-1)d},  \\
     =\begin{cases}
        \tau^{2\alpha\gamma},&d=3,\\
        1,&\forall\; d>3,
     \end{cases}
\end{split}
\end{equation}
can be shown.
In the $d=3$ case, we use
\begin{equation}
 \label{eq3.2.7}
  C_{\left(\begin{bmatrix}
1&0\\
\gamma&1
\end{bmatrix} \middle| \begin{bmatrix}
x\\
z
\end{bmatrix}\right)} \in SU(p)\;\forall\; p>3
\end{equation}
and
\begin{equation} 
\label{eq3.2.8}
   \det\left( C_{\left(\begin{bmatrix}
1&0\\
\gamma&1
\end{bmatrix} \middle| \begin{bmatrix}
x\\
z
\end{bmatrix}\right)}\right) =\tau^{2\gamma}\; \text{for}\; p=3. 
\end{equation}

With all the mathematical definitions at hand, we are ready to give an explicit form of the qudit~$\pi/8$ gate. We choose the qudit gate $U_{\upsilon}$ to be diagonal in the computational basis and claim that,
for $d>3$,
$U_{\upsilon}$ has the form 
\begin{equation}
\label{eq3.2.3}
U_{\upsilon}=U(\upsilon_0,\upsilon_1,\ldots)=\sum^{d-1}_{k=0}\omega^{\upsilon_k}\ket{k}\bra{k}(\upsilon_k\in\mathbb{Z}_d).
\end{equation}
A straightforward application of Eqs.~(\ref{eq:displacement})
and~(\ref{eq3.2.3}) yields
\begin{equation}
\label{eq3.2.4}
U_{\upsilon}D_{(x|z)}U_{\upsilon}^\dagger=D_{(x|z)}\sum_k \omega^{\upsilon_{k+1}
-\upsilon_k}\ket{k}\bra{k}.
\end{equation}
As $U_{\upsilon}$ is to be a member of $\mathcal{C}_3$, the right hand side of Eq.~(\ref{eq3.2.4})
must be a Clifford gate.
We ignore the trivial case $U_{\upsilon}D_{(0|z)}U_{\upsilon}^\dagger=D_{(0|z)}$ and focus on the case $U_{\upsilon}D_{(1|0)}U_{\upsilon}^\dagger$ in order to derive an explicit expression for $U_{\upsilon}$.

We define $\gamma',z',\epsilon'\in\mathbb{Z}_d$ such that 
\begin{equation}
\label{eq3.2.5}
U_{\upsilon}D_{(1|0)}U_{\upsilon}^\dagger
=\omega^{\epsilon'}C_{\left(\begin{bmatrix}
1&0\\
\gamma'&1
\end{bmatrix} \middle| \begin{bmatrix}
1\\
z'
\end{bmatrix}\right)}
\end{equation}
From Eqs.~(\ref{eq3.2.2}) and~(\ref{eq3.2.4})
we see that the right-hand side of Eq.~(\ref{eq3.2.5}) is the most general form, and we note that $U\in SU(d)$ implies $\omega^{\gamma'}U\in SU(d)$.
We rewrite the left-hand side of Eq.~(\ref{eq3.2.5}) using Eq.~(\ref{eq3.2.4}) and right-hand side using Eq.~(\ref{eq3.2.2}) and obtain
\begin{equation}
    D_{(1|0)}\sum_k \omega^{\upsilon_{k+1}-\upsilon_k}\ket{k}\bra{k}=\omega^{\epsilon'}D_{(1|z')}\sum^{d-1}_{k=0}\tau^{\gamma'k^2}\ket{k}\bra{k}.
\end{equation}
After cancelling common factors of $D_{(1|0)}$,
an identity between two diagonal matrices remains such that 
\begin{equation}
    \omega^{\upsilon_{k+1}-\upsilon_k}=\omega^{\epsilon'}\tau^{z'}\omega^{kz'}\tau^{\gamma'k^2} \;(\forall k\in \mathbb{Z}_d),
\end{equation}
or, equivalently, using Eq.~(\ref{eq:displacement}),
\begin{equation}
\upsilon_{k+1}-\upsilon_k=\epsilon'+2^{-1}z'+kz'+2^{-1}\gamma'k^2.
\end{equation}
From here, we derive the recursive relation
\begin{equation}
\upsilon_{k+1}=\upsilon_k+k(2^{-1}\gamma'k+z')+2^{-1}z'+\epsilon'.
\end{equation}
We solve for the $\upsilon_k$ with a boundary condition $\upsilon_{0}=0, $
\begin{equation}
\upsilon_k=\frac{1}{12}k\{\gamma'+k[6z'+(2k-3)\gamma']\}+k\epsilon',
\end{equation}
where all factors are evaluated modulo~$d$. For example, with $d=5$, the fifth root of unity~(\ref{eq:omega}) is $\omega=\text{e}^{2\pi\text{i}/5}$ and choosing $z'=1,\gamma'=4$ and $\epsilon'=0$,
we obtain 
\begin{equation}
\label{eq:upsilon01234}
\upsilon=(\upsilon_0,\upsilon_1,\upsilon_2,\upsilon_3,\upsilon_4)=(0,3,4,2,1)
\end{equation}
so that
\begin{equation}
    U_{\upsilon}=\begin{pmatrix}
\omega^0&0&0&0&0\\
0&\omega^{-2}&0&0&0\\
0&0&\omega^{-1}&0&0\\
0&0&0&\omega^{2}&0\\
0&0&0&0&\omega^{1}
\end{pmatrix}
\end{equation}
The diagonal elements of $U_{\upsilon}$ are powers of $\omega$ that sum to zero modulo~$d$ and, consequently,
$\operatorname{det}(U_{\upsilon})=1.$ 

For the $d=3$ case, because of Eq.~(\ref{eq3.2.6}) extra work is needed for solving a matrix equation similar to Eq.~(\ref{eq3.2.5}).
We first introduce a global phase factor
$\text{e}^{\text{i}\phi}$  such that
\begin{equation}
\label{eq3.2.9}
    \det\left(\text{e}^{\text{i}\phi}\sum^{d-1}_{k=0}\tau^{\gamma k^2}| k\rangle\bra{k}\right)=1\implies\phi=4\pi\gamma/9.
\end{equation}
The ninth root of unity~(\ref{eq:omega}) is $\omega=\text{e}^{2\pi\text{i}/9}$ and, from Eq.~(\ref{eq3.2.8}) we derive that
\begin{equation}
    \det\left( \omega^{2\gamma'}C_{\left(\begin{bmatrix}
1&0\\
\gamma'&1
\end{bmatrix} \middle| \begin{bmatrix}
1\\
z'
\end{bmatrix}\right)}\right) =1.
\end{equation}
The qutrit version of~$U_{\pi/8}$ has a more general form than in Eq.~(\ref{eq3.2.3});
i.e.
\begin{equation}
    U_{\upsilon}=U(\upsilon_0,\upsilon_1,\ldots)
    =\sum^{2}_{k=0}\omega^{\upsilon_k}\ket{k}\bra{k},\;
    \upsilon_k\in\mathbb{Z}_9.
\end{equation}
Then the general solution is
\begin{equation}
    \upsilon=(0,6z'+2\gamma'+3\epsilon',6z'+\gamma'+6\epsilon')\mod 9.
\end{equation}
For example, choosing $z'=1,\gamma'=2$ and $\epsilon'=0$,
\begin{equation}
    U_{\upsilon}=\begin{pmatrix}
    \omega^{0}&0&0\\
    0&\omega^{1}&0\\
    0&0 &\omega^{-1}
    \end{pmatrix}.
\end{equation}

The~$\pi/8$ gate, with its close relation to the Clifford group, has many applications and utilities in teleportation-based UQC~\cite{Gottesman1999}, transversal implementation~\cite{Eastin2009,Zeng2007}, learning an unknown gate~\cite{Low2009}, or securing assisted quantum computation~\cite{Childs2001}.
The generalized qudit version of the~$\pi/8$ gate, $U_{\upsilon}$, is shown to be identical to the maximally robust qudit gates for qudit fault-tolerant UQC discussed in reference~\cite{vanDam2011}.

This gate also plays an important role in the magic-state distillation (MSD) protocols for general qudit systems,
which was first established for qutrits~\cite{ACB12}
and then extended to all prime-dimensional qudits~\cite{CAB12}.
\subsubsection{Qudit SWAP gate}
\label{subsubsec:swap}
\begin{figure}[h!]
\begin{center}
\includegraphics[width=15cm]{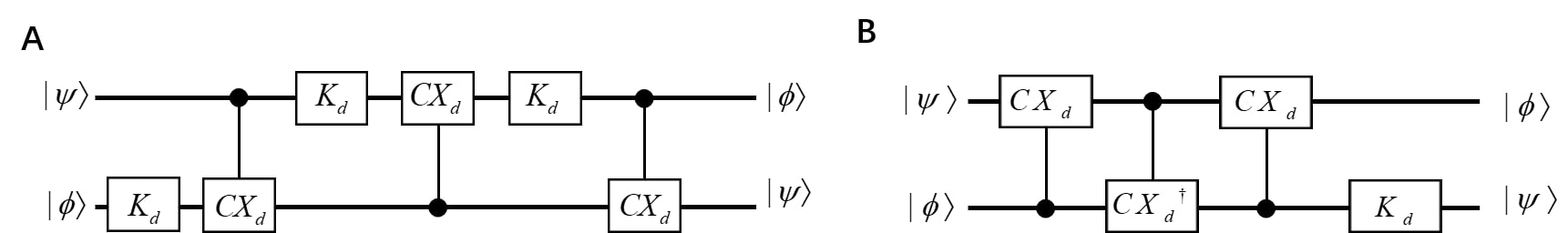}\\
\caption{\textbf{(A)} is the qudit SWAP circuit using $CX_d$ and $K_d$ gates~\cite{Fujii_2003,Paz-Silva2009}. \textbf{(B)} is the qudit SWAP circuits with the $CX_d$, the $CX_d^\dagger$ and the $K_d$ gates } 
\label{fig:qudit_SWAP2}
\end{center}
\end{figure}

A SWAP gate is used to exchange the states of two qudit such that:
\begin{equation}
\mathsf{SWAP}|\phi\rangle|\psi\rangle=|\psi\rangle|\phi\rangle
\end{equation}
Various methods to achieve the SWAP gate use different variants of qudit controlled gates~\cite{WILMOTT2011,WILMOTT2012,Mermin2001,Fujii_2003,Paz-Silva2009,Wang_2001,Alber_2001} as shown in Fig.\ref{fig:qudit_SWAP2}. The most used component of the SWAP gate is a controlled-shift gate $CX_d$ that perform the following operation:
\begin{equation}
    CX_d\ket{x}\ket{y}=\ket{x}|x+y\rangle
\end{equation}
with a modulo~$d$ addition.
Its inverse operation is
\begin{equation}
    CX_d^\dagger\ket{x}\ket{y}=\ket{x}|y-x\rangle
\end{equation}
In some approaches,
the operation $K_d$ is required to complete the circuits, where 
\begin{equation}
    K_d\ket{x}=|d-x\rangle=\ket{-x},
\end{equation}
which outputs the modulo~$d$ complement of the input.
These circuits are more complex and less intuitive then the qubit SWAP gate~\cite{Fujii_2003} because they are not Hermitian,
i.e., $CX_d\neq CX_d^\dagger$.

\begin{figure}[h!]
\begin{center}
\includegraphics[width=10cm]{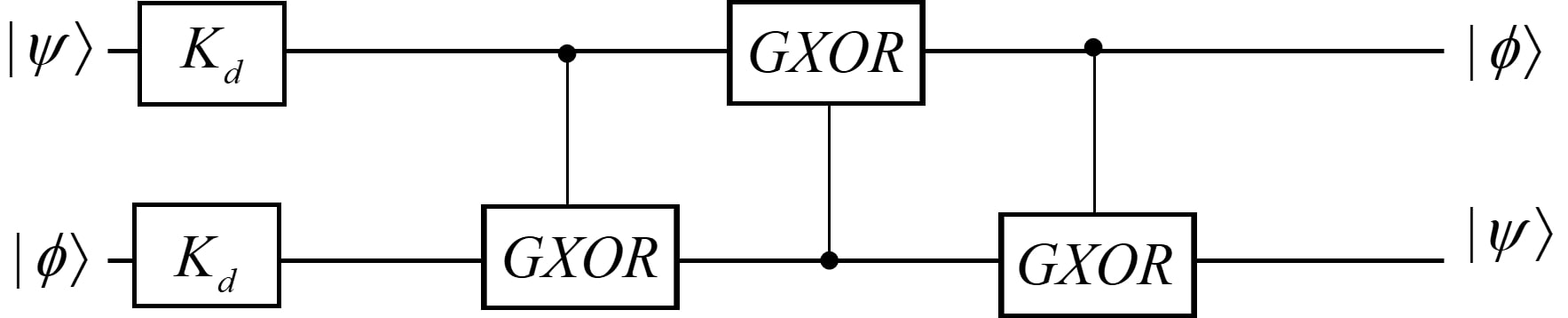}\\
\caption{Qudit SWAP circuits with the $GXOR$ and the $K_d$ gates~\cite{Wang_2001,Alber_2001}.}
\label{fig:qudit_SWAP4}
\end{center}
\end{figure}

One way to create a Hermitian version of the qudit CNOT uses the GXOR gate
\begin{equation}
    GXOR\ket{x}\ket{y}=\ket{x}\ket{x-y}.
\end{equation}
However, this SWAP gate needs to be corrected with an $K_d$~\cite{Alber_2001} as shown in Fig.~\ref{fig:qudit_SWAP4}.
A partial SWAP gate $S_p$~\cite{DWS03}
works on a hybrid system where~$\ket{i}$ is a qudit of dimension $d_c$ and~$\ket{j}$ is a qudit of dimension $d_t$
\begin{equation}
    S_p\ket{i}\otimes\ket{j}=\begin{cases}
    \ket{j}\otimes\ket{i} &\;\; \text{for}\; i,j\in \mathbb{Z}_{d_p} \\
    \ket{i}\otimes\ket{j} &\;\; \mathrm{otherwise}
    \end{cases}
\end{equation}
where $d_p\leqslant d_\text{min}=\min(d_c,d_t)$

\begin{figure}[h!]
\begin{center}
\includegraphics[width=15cm]{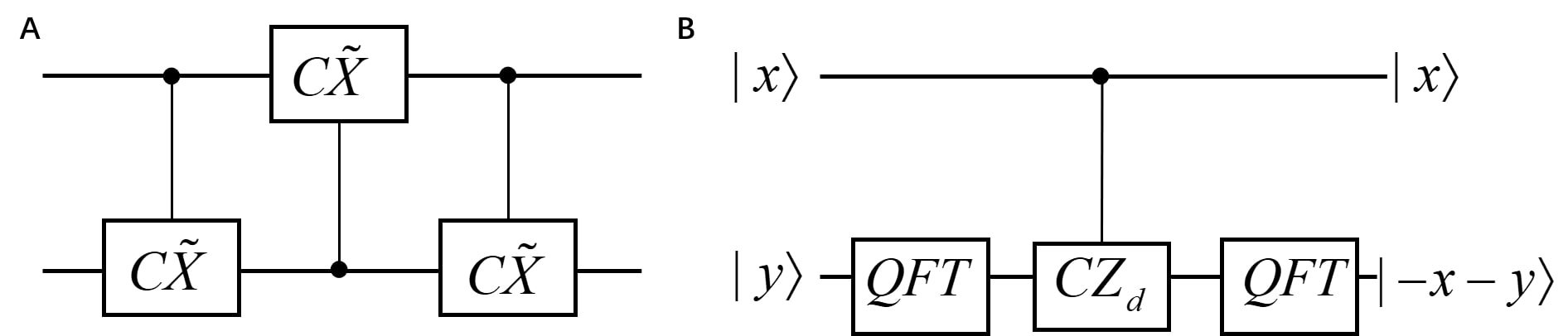}\\
\caption{\textbf{(A)} is the qudit SWAP gate with the $C\tilde{X}$ gate. \textbf{(B)} is the decomposing $C\tilde{X}$ gate. The $QFT$ represents the quantum Fourier transform while $CZ_d$ is the selective phase shift gate.} 
\label{fig:qudit_SWAP}
\end{center}
\end{figure}

In the rest of this section, we present a  Hermitian generalization of the qudit CNOT gate with a symmetry configuration and  a qudit SWAP circuit with a single type of qudit gate as shown in Fig.~\ref{fig:qudit_SWAP} \textbf{A}~\cite{Garcia-Escartin2013}. Compared with all the previously proposed SWAP gate for qudit, this method is easier to implement since there is only one type of gate $C\tilde{X}$ needed. To begin with, we define a gate $C\tilde{X}$ acting on~$d$-level qudits $\ket{x}$ and $\ket{y}$ such that 
\begin{equation}
C\tilde{X}\ket{x}\ket{y}=\ket{x}\ket{-x-y},
\end{equation}
where $\ket{-x-y}$ represents a state $|i=-x-y\rangle$ in the range $i\in\{0,\ldots,d-1\}$ mod~$d$.
Notice that,
for $d=2$, the $C\tilde{X}$ gate is equivalent to the CNOT gate.
The SWAP gate for qudit can be built using three  $C\tilde{X}$ gates.

$C\tilde{X}$ is generated with three steps: a qudit generalization of the $CZ$ gate as $CZ_d$ sandwiched by two quantum Fourier transform operations(QFT).  The circuit illustration for the sequence of theses gate is shown in Fig.~\ref{fig:qudit_SWAP} \textbf{B}.
The QFT transforms the~$\ket{x}$ into a uniform superposition
\begin{equation}
QFT\ket{x}=\frac{1}{\sqrt{d}}\sum_{k=0}^{d-1}\text{e}^{\text{i}2\pi xk/d}\ket{k}.
\end{equation}
The $CZ_d$ gate adds a phase to the target qudit depending on the state of the control qudit. Its effect on the input qudits is
\begin{equation}
CZ_d\ket{x}\ket{y}=\text{e}^{\text{i}2\pi xy/d}\ket{x}\ket{y}.
\end{equation} 
The inverse QFT undoes the Fourier transform process and the inverse of $CZ_d$ is
\begin{equation}
CZ^\dagger_d\ket{x}\ket{y}=\text{e}^{-\text{i}2\pi xy/d}\ket{x}\ket{y}.
\end{equation}
The full evolution of the $C\tilde{X}$ is 
\begin{align}
\ket{x}\ket{y}  &\xrightarrow{QFT_2}\frac{1}{\sqrt{d}}\sum^{d-1}_{k=0}\text{e}^{\text{i}\frac{2\pi ky}{d}}\ket{x}\ket{k}\\
 &\xrightarrow{CZ_d}\frac{1}{\sqrt{d}}\sum^{d-1}_{k=0}\text{e}^{\text{i}\frac{2\pi ky}{d}}\text{e}^{\text{i}\frac{2\pi xk}{d}}\ket{x}\ket{k}=\frac{1}{\sqrt{d}}\sum^{d-1}_{k=0}\text{e}^{\text{i}\frac{2\pi k(x+y)}{d}}\ket{x}\ket{k}\\
 &\xrightarrow{QFT_2}\frac{1}{d}\sum^{d-1}_{l=0}\sum^{d-1}_{k=0}\text{e}^{\text{i}\frac{2\pi k(x+y)}{d}}\text{e}^{\text{i}\frac{2\pi kl}{d}}\ket{x}|l\rangle=\ket{x}\ket{-x-y}.
\end{align}
It is easy to show that $C\tilde{X}$  is its own inverse and then $C\tilde{X}=C\tilde{X}^\dagger$.
For the proposed SWAP gate, both the QFT and $CZ_d$ operations are realizable on a multilevel quantum systems. For example, there
are implementations of them for multilevel atoms~\cite{Muthukrishnan2000,Muthukrishnan2002}. The resulting SWAP gate provides a way to connect systems limited to the nearest-neighbour interactions. 
This gate provides a useful tool in the design and analysis of complex qudit circuits.
\subsubsection{Simplified qubit Toffoli gate with a qudit}
\label{sec:improv_toffoli_gate}
\begin{figure}[h!]
\begin{center}
\includegraphics[width=10cm]{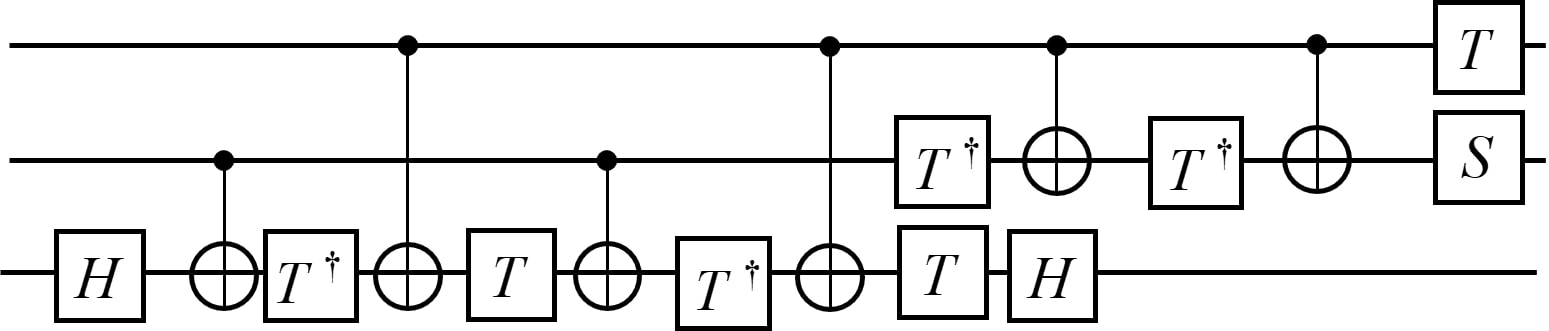}\\
\caption{Decomposing qubit Toffoli gate with the universal qubit gates. $H$ is the Hadamard gate, $T$ is the $\pi / 8$ gate and $S$ is the phase gate.} 
\label{fig:qubit_toffoli}
\end{center}
\end{figure}

The Toffoli gate is well known for its application to universal reversible classical computation.
In the field of quantum computing, the Toffoli gate plays a central role in quantum error correction~\cite{Cory1998},
fault tolerance~\cite{Dennis2001} and offers a simple universal quantum gate set combined with one qubit Hadamard gates~\cite{shi2002}.
The simplest known qubit Toffoli gate, shown in Fig.~\ref{fig:qubit_toffoli}, requires at least $5$ two-qubit gates~\cite{Nielsen2011}. However, if the target qubit has a third level, i.e., a qutrit, the whole circuit can be achieved with three two-qubit gates~\cite{Ralph2007}.

\begin{figure}[h!]
\begin{center}
\includegraphics[width=10cm]{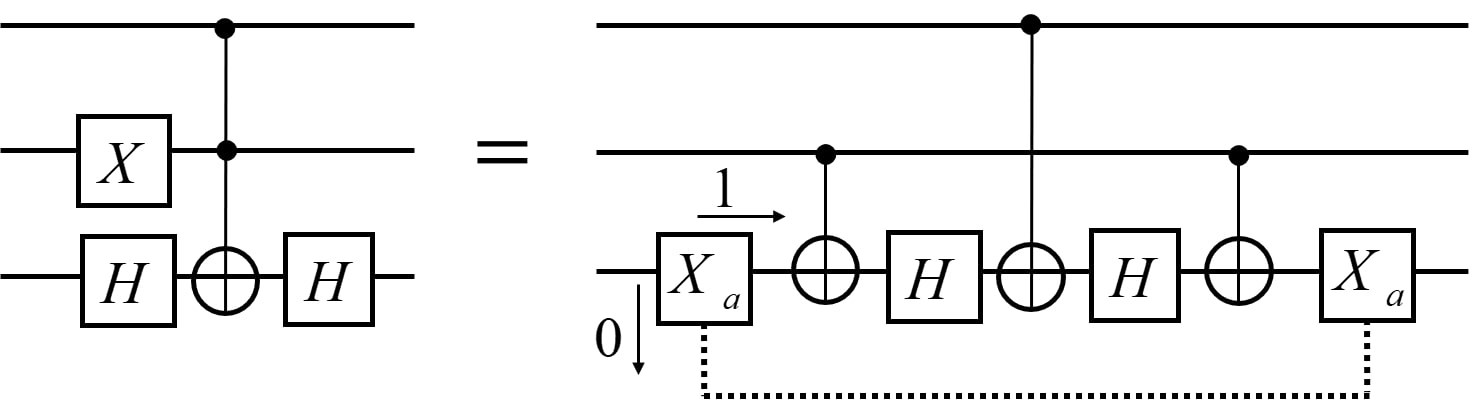}\\
\caption{The Simplified Toffoli gate. The first two lines represent two control qubits and the third line represents a target qutrit that has three accessible levels. 
The initial and final quantum states of the quantum information carrier are encoded in the $\ket0$ and $\ket1$.
The $H$ is the  generalized Hadamard gate such that $H\ket0=1/\sqrt2[\ket0$+$\ket1]$, $H\ket1=1/\sqrt2[\ket0$ - $\ket1]$ and $H\ket{2}=\ket{2}$.  $X_a$ gate is a qutrit gate such that $X_a\ket0=\ket{2}$ and $X_a\ket{2}=\ket0$ with  $X_a\ket1=\ket1$. With the control being qubit, the target being qudit, the two qudit gate in this case is a hybrid gate. } 
\label{fig:qudit_toffoli}
\end{center}
\end{figure}

A new qutrit gate $X_a$ is introduced to the circuit that does the following: $X_a\ket0=\ket{2}$ and $X_a\ket{2}=\ket0$ with  $X_a\ket1=\ket1$.
The simplified circuit is shown in Fig.~\ref{fig:qudit_toffoli}.
The two controlled gates are the CNOT gate and a control-$Z$ gate,
which is achieved with a CNOT gate between two Hadamard gates.
The Hadamard gate here operating on the qutrit is generalized from the normal Hadamard gate operating on a qubit---it only works with the $\ket0$ and $\ket1$, such that $H\ket0=1/\sqrt2[\ket0$+$\ket1]$, $H\ket1=1/\sqrt2[\ket0$ - $\ket1]$ and $H\ket{2}=\ket{2}$. 
Comparing the circuit in Fig.~\ref{fig:qudit_toffoli}  to that in Fig.~\ref{fig:qubit_toffoli}, it is clear that the total number of gates is significantly reduced.

This method can be generalized to $n$-qubit-controlled Toffoli gates by utilizing a
single ($n+1$)-level target carrier and using only $2n-1$ two-qubit gates~\cite{Ralph2007}. In other words, the target carrier needs an extra level for each extra control qubit. Compare to the best known realization previously that requires $12n-11$ two-qubit gates~\cite{Nielsen2011},
this method offers a significant resource reduction.
Furthermore, these schemes can be extended to more general quantum circuits such as the multi-qudit-controlled-unitary gate $C^nU$. 

The previous method turns the target qubit into a qudit;
another method simplifies the Toffoli gate by using only qudits and treating the first two levels of the qudit as qubit levels and other levels as auxiliary levels.
The reduction in the complexity of Toffoli gate is accomplished by utilizing the topological relations between the dimensionality of the qudits, where higher qudit levels serve as the ancillas~\cite{KNX+20}.

Suppose we have a system of~$n$ qudits denoted as $Q_i,\;i\in \{1,\ldots,n\}$ and each qudit has dimension $d_i\geqslant 2$.
Qudits are initialized into pure or mix states on the first two levels,
i.e., the qubit states,
and zero population for the other levels,
i.e., the auxiliary states.
This scheme assumes the ability to perform single-qubit operations. We can apply the desirable unitary operation on the qubit states and leave the auxiliary states unchanged. We also assume that we have the ability  to manipulate the auxiliary levels by a generalized inverting gate $X_m$
\begin{equation}
    X_m\ket0=\ket{m},\: X_m\ket{m}=\ket0,\: X_m\ket{y}=\ket{y},\, \text{for}\,y\neq m,0.
\end{equation}
At the same time, the two-qubit $CZ$ gates are applied according to certain topological connections between qudits. We introduce a set $E$ of ordered pairs $(i,j)$, such that $i,j\in\{1,\ldots,n\},\, i<j$ to obtain this topology and the $CZ$ gate is defined as
\begin{equation}
    CZ\ket{11}_{Q_i,Q_j}=-\ket{11}_{Q_i,Q_j}\;CZ\ket{xy}_{Q_i,Q_j}=\ket{xy}_{Q_i,Q_j}\:\text{for}\,xy\neq 1,
\end{equation}
with $x\in\{0,\ldots d_i-1\}$ and $y\in\{0,\ldots d_j-1\}$. 

The set $E$ describes an $n$-vertex-connected
graph. Let $\tilde{E}\subseteq E$ defines an $n$-vertex connected \textit{tree} (acyclic graph). The main result is:
the $n$-qubit Toffoli gate can be achieved with less number of operations if
\begin{equation}
    d_i\geqslant k_i +1,
\end{equation}
where $d_i$ is the dimension of a qudit and the number $k_i$ is the qudit's connections to other qudits
within $\tilde{E}$. With this condition fulfilled, the $n$-qubit Toffoli gate can be realized by $2n-3$ two-qudit $CZ$ gates. The detailed realization of the $n$-qubit Toffoli gate by the properties and special operations of the tree in topology can be found in Ref.~\cite{KNX+20}.
The advantage of this scheme is the scalability and the ability to implement it for the multi-qubit controlled unitary gate $C^{n}U$.

These $C^nU$ gates are a crucial component in the PEA which has many important applications such as the quantum simulation~\cite{Aspuru-Guzik1704} and Shor's factoring algorithm~\cite{shor1994}. This idea of combining qudits of different dimensions or hybrid qudit gates can also be applied to other qudit gates such as the SWAP and SUM gates as shown in Refs.~\cite{DWS03,CEMM98}. Thus, introducing qudits into qubit systems to create a hybrid qudit system offers the potential of improvement to quantum computation.

\subsubsection{Qudit multi-level controlled gate}
\label{sec:qudit_MVCG}

For a qubit controlled gate, the control qubit has only two states so it is a ``do-or-don't'' gate. Qudits, on the other hand, have multiple accessible states and thus a qudit-controlled gate can perform a more complicated operation~\cite{Di2013}. The \textit{Muthukrishan-Stroud gate} (MS gate) for a qudit applies the specified operation on the target qudit only if the control qudit is in a selected one of the $d$ states, and leaves the target unchanged if the control qudit is in any other $d-1$ states.
Hence, the MS gate is essentially a "do-or-don't" gate generalized to qudits and does not fully utilize the~$d$ states on the control qudit~\cite{Muthukrishnan2000}.

\begin{figure}[h!]
\begin{center}
\includegraphics[width=10cm]{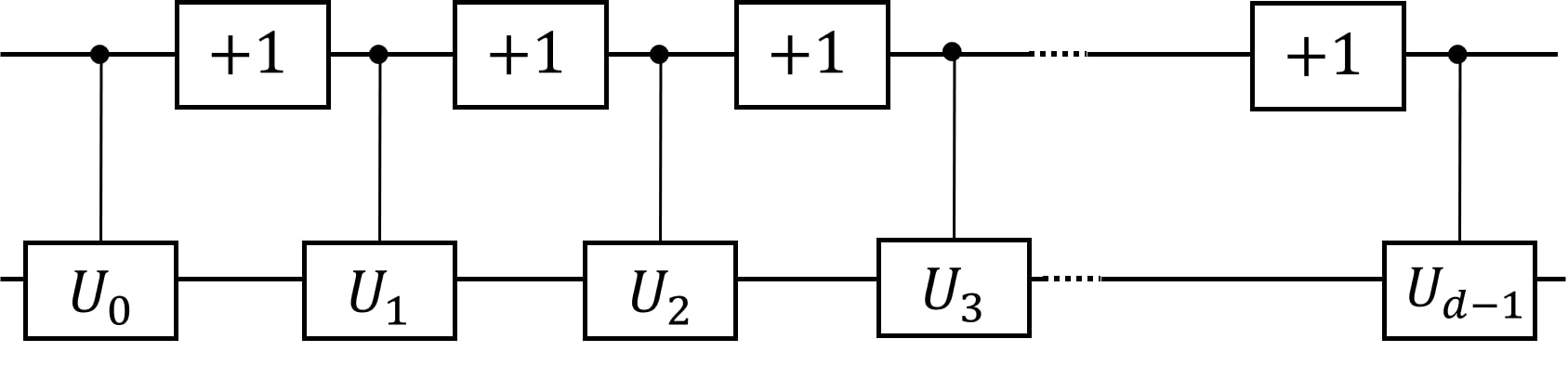}\\
\caption{$d$-valued Quantum Multiplexer for the second qutrit and its realization in terms of Muthukrishan-Stroud gates(the control~$U$ operation that only act on one specific control state). The gate labeled $+1$ is the shifting gate that increases the state value of the control qudit by $1$(mod~$d$). Depending on
the value of the top control qudit, one of $U_i$ is applied to the second qudit, for $i\in \{0,1,\ldots d-1\}$. } 
\label{fig:qudit_MS}
\end{center}
\end{figure}

To fully utilize the~$d$ states on the control qudit, people have developed the quantum multiplexer to perform the controlled~$U$ operations in a qudit system as shown in Fig.~\ref{fig:qudit_MS}, where the MS gate and shifting gates are combined to apply different operations to the target depending on different states on the control states~\cite{KHAN2006336}. Here we discuss the \textit{multi-value-controlled gate} ($MVCG$) for qudits, which applies a unique operation to the target qudit for each unique state of the control qudit~\cite{Hsuan-Hao2019}.

For a~$d$-dimensional qudit system, a two-qudit multi-value-controlled gate is represented by a $d^2\times d^2$ matrix
\begin{equation}
MVCG=\begin{pmatrix}
    U_0 & 0 & 0 & \cdots & 0  \\
    0 & U_1 & 0 & \cdots & 0  \\
    0 & 0 & U_2 & \cdots & 0  \\
    \vdots & \vdots & \vdots & \ddots & 0  \\
    0 & 0 & 0 & \cdots & U_{d-1}
    \end{pmatrix},
\end{equation}
where each $U_i$ ($i=0,1,\ldots, d-1 $) is a unique unitary single-qudit operation. The $U_i$ operation is applied to the target qudit when the control qudit is in~$\ket{i}$ state.  In the later sections, \S\ref{sec:qudit QFT} and \S\ref{sec:qudit PEA} the controlled gates are $MVCG$ and improve the efficiency of the qudit algorithm.
$MVCG$ can be built in many physical systems and one example in a photonic system  is introduced in \S\ref{subsec:timefreqphoton}.

\subsection{Geometrically quantifying qudit-gate efficiency }
\label{sec:gategeometry}
 In a quantum computer, each qudit can remain coherent for a limited amount of time (decoherence time). After this time, the quantum information is lost due to the outside perturbations and noises. In the computation process, quantum gates take certain amount of time to alter the states of the qudits.
The decoherence time of a qudit state limits the number of quantum gates in the circuit. Therefore, we need to design  more efficient algorithms and circuits. 
A method exists to do a general systematic evaluation of the circuit efficiency with the mathematical techniques of Riemannian geometry~\cite{nielsen2006quantum}. By reforming the 
quantum circuits designing problems as a geometric problem, we are able to develop new quantum algorithms or to exploring and evaluating the full potential of the quantum computers.
This evaluation is able to generalized to qutrit  systems, where the least amount of the gates required to synthesize any unitary operation is given~\cite{li2013qutrits}.
 
To begin with,
we assume that the operations done by the quantum circuit can be described by a unitary evolution~$U$ derived from the time-dependent Schr\"odinger equation $\text{d}U/\text{d}t=-\text{i}HU$ with the boundary condition~$t_\text{f}$, $U(t_\text{f})=U$. The complexity of realizing $U$ can be characterized by a cost function $F[H(t)]$ on the Hamiltonian control $H(t)$. This allow us to define a Riemannian geometry on the space of unitary operations~\cite{nielsen2005geometric}.
Finding the minimal geodesics of this Riemannian geometry is equivalent to finding the optimal control function~$H(t)$ of synthesizing the desired $U$.

Now we transform the problem of calculating a lower bound to the gate number to finding the minimal geodesic distance  between the identity operation~$I$ and~$U$.
Instead of Pauli matrices for the qubit representation of the Hamiltonian, the qutrit version of Hamiltonian is expanded in terms of the Gell-Mann matrices.
Here we give an explicit form of the Gell-Mann matrices representation in~$d$-dimension~\cite{luo2014qudits} which is used for qutrit (where $d=3$) as well as other qudit systems in the later part of the section. Let $e_{jk}$ denote the $d\times d$ matrix with a $1$ in the ($j,k$) elements and $0$s elsewhere, a basis can be described as
\begin{align}
     u^d_{jk}&=e_{jk}+e_{kj},\; 1\leq j<k\leq d,\\
      u^d_{jk}&=\text{i}(e_{jk}-e_{kj}),\; 1\leq k<j\leq d,\\
      u^d_{jj}&=\text{diag}(1,\ldots,1,-j,0_{d-2j}),j\in[d-1].
\end{align}
Here, diag represents the diagonal matrix,
$0_{d-2j}$ denotes the zeros of length $d-2j$.
$u^d_{jk}$ are traceless and Hermitian and together with the identity matrix $\mathds1_d$ serve as the basis of the vector space of $d\times d$ Hermitian matrix.
These generalized Gell-Mann matrices can be used to generate the group representation of $SU(d)$ while the other representations can be achieved by transform these matrices uniformly. To derive the bases of $SU(d^n)$, we first define $x_l=u^d_{jk}$ with $l=jd+k,l\in\left[d^2\right]$ and 
\begin{equation}
    X^s_{l}=\text{I}^{\otimes s-1}\otimes x_l\otimes I^{\otimes  d-s}
\end{equation}
acts on the $s$-th qudit with $x_l$ and leaves the other qudits unchanged. The bases of  of $SU(d^n)$ is constructed by $\{Y^{P_t}_t\},t\in\left[n\right],
 P_t=\{i_1,\ldots,i_t\}$ with all possible $1<i_1<\cdots<i_k<n$, where
 \begin{equation}
     Y^{P_t}_t=\prod^t_{k=1}X^{i_k}_{j_k}.
 \end{equation}
 $Y^P_t$ denotes all operators with generalized Gell-Mann matrices $x_{j_1},\ldots,x_{j_k}$ acting on $t$ qudits at sites $P=\{i_1,\ldots,i_k\}$, respectively, and rest with identity. It is easy to prove that with the generalized Gell-Mann matrices representations,1-body and 2-body interactions can generate all 3-body interactions. 
 
Now the Hamiltonian in terms of the Gell-Mann matries (with the notation $\sigma$) can be written as
 \begin{equation}
 \label{eq:HPauliexpansion}    
    H=\sum'_{\sigma}h_{\sigma}\sigma+\sum^{\prime\prime}_{\sigma}h_{\sigma}\sigma.
 \end{equation}
All coefficients~$h_{\sigma}$ are real and,
in $\sum'_{\sigma}h_{\sigma}\sigma$,
$\sigma$ goes over all possible one- and two-body interactions whereas,
in $\sum^{\prime\prime}_{\sigma}h_{\sigma}\sigma$,
$\sigma$ goes over everything else.
The cost function is
\begin{equation}
    F(H):=\sqrt{\sum'_{\sigma}h^2_{\sigma}\sigma+p^2\sum^{\prime\prime}_{\sigma}h^2_{\sigma}\sigma},
\end{equation}
 where~$p$ is a penalty cost by applying many-body terms. Now that the control cost is well defined, it is natural  to form the distance in the space $SU(3^n)$ of $n$-qutrit unitary operators with unit determinant. We can treat the function $F(H)$ as the norm related to a  Riemannian metric with a metric tensor $g$ as:
 \begin{equation}
 \label{eq. metric}
     g=\begin{cases}
     0,&\sigma\neq \tau \\
     1,&\sigma=\tau$ and $\sigma $ is one or two body $\\
     p^2,&\sigma=\tau$ and $\sigma $ is three or more body $
     \end{cases}.
 \end{equation}
 The distance $d(I, U)$ between~$I$ and~$U$  which is the minimum curve connecting~$I$ and~$U$ equals to the minimal length solution to the geodesic equation
 \begin{equation}
\left\langle\frac{\text{d}H}{\text{d}t},K \right\rangle
    =\text{i}\left\langle H,[H,K]\right\rangle,
 \end{equation}
 where $\langle,\rangle$ denotes the inner product on the tangent space $SU(3^n)$ defined by the metric components~(\ref{eq. metric}),
 and $[,]$ denotes the matrix commutator and~$K$ is an arbitrary operator in $SU(3^n)$.

All lemmas backing up the final theorem have been proven in detail~\cite{li2013qutrits},
but the reasoning behind can be summarized in four parts.
First let~$p$ be the three- and more-body
items penalty.
With large enough $p$, the distance $d(I,U)$ is guarantee to have a  supremum that does not depend on $p$.
Secondly, we have
\begin{equation}
    \|U-U_P\| \leqslant 3^n d([U])/p,
\end{equation}
where $\|\bullet\|$ is the operator norm and $U_P$ the corresponding unitary operator generated by the one- and two-body items projected Hamiltonian $H_P(t)$.
Thirdly, given an $n$-qutrit unitary operator $U$ generated by $H(t)$ with the condition $\|H(t)\|\leqslant c$ in a time interval $[0,\Delta]$,
then
\begin{equation}
    \|U-\exp(-\text{i}\bar{H})\Delta\|\leqslant 2(\text{e}^{c\Delta}-1-c\Delta)
    =O(c^2\Delta^2),
\end{equation}
where $\bar{H}$ is the mean Hamiltonian. Lastly, for $H$ as an $n$-qutrit one- and two-body Hamiltonian,
a unitary operator $U_A$ exists that satisfies \begin{equation}
    \|\text{e}^{\text{i}H\Delta}-U_A\|\leqslant c_2n^2\Delta^3
\end{equation}
and can be generated with at most $c_1n^2/\Delta $ one- and two-qutrit gates, and constants $c_1$ and $c_2$.

All these lemmas combined gives the final theorem for the qutrit system:
for a unitary operator~$U$ in $SU(3^n)$, $O(n^kd(I, U)^3)$  one- and two-qutrit gates is the lower bound  to synthesize a unitary $U_A$ with the condition $\Vert U-U_A \Vert\leq c$, given a constant $c$. It is worth mentioning that for any groups of unitaries $U$, which is labeled by the number of qudits $n$,
the final theorem shows a quantum circuit exists with a  polynomial of $d(I, U)$ number of gates such that it can approximates $U$ to arbitrary accuracy. Alternatively,a polynomial-sized quantum circuit exists if and only if the distance $d(I, U)$ itself is scaling polynomially with $n$.

With appropriate modification,
the Riemannian geometry method can be used to ascertain the circuit-complexity bound for a qudit system~\cite{luo2014qudits}.
In this scheme, the unitary matrix $U\in SU(d^n)$ is represented by the generalized Gell-Mann matrices as defined in the earlier part of the section.
The main theorem in the qudit case of the Ref.~\cite{luo2014qudits} is `` for any small constant $\varepsilon$, each unitary $U_A\in SU(d^n)$ can be synthesized using $O(\varepsilon^{-2})$ one- and two-qudit gates, with error $\Vert U-U_A\Vert\leq\varepsilon$.'' To break up the constant $\varepsilon$ to an explicit form, we have $\varepsilon^{-2}=N^2d^4n^2$,
where~$d$ is the dimension of the qudit,
$n$ is the number of qudits and~$N$ is the number of the intervals that~$d(I, U)$ divides into, such that a small $\delta=d(I,U)/N\leqslant \varepsilon$. The qudit case shows the explicit relation between the non-local quantum gate cost and the approximation error for synthesizing quantum qudit operations. In summary, for the quantum circuit model, one can decide a lower bound for the number of gates needed to synthesize~$U$ by finding the shortest geodesic curve linking~$I$ and~$U$. This provides a good reference for the design of the quantum circuit using qudits.

\section{Quantum algorithms using qudits}
\label{sec:qudit alg}
A qudit, with its multi-dimensional nature, is able to store and process a larger amount of information than a qubit.  Some of the algorithms described in this section can be treated as direct generalizations of their qubit counterparts and some utilize the multi-dimensional nature of the qudit at the key subroutine of the process. This section introduces examples of the well-known quantum algorithms based on qudits and divides them into two groups: algorithms for the oracle-decision problems in~\S\ref{sec:quditoracledecision} and algorithms for the hidden Abelian subgroup problems in~\S\ref{sec:qudithiddenabelian}. Finally,  \S\ref{sec:qudti search} discusses how the qudit gates can improve the efficiency of the quantum search algorithm and reduce the difficulty in its physical set-up. 

\subsection{Qudit oracle-decision algorithm}
\label{sec:quditoracledecision}
In this subsection we explore the qudit generalizations of the efficient algorithms for solving the oracle decision problems,
which are quite important historically and used to demonstrate the classical-quantum complexity separation~\cite{Deu85,DJ92}.
The oracle decision problems is
to locate the contents we want from one of the two mutually disjoint sets that is given. 
We start in \S\ref{sec:speedup alg} with a discussion about a single-qudit algorithm that determines the parity of a permutation.
In~\S\ref{subsubsec:quditDJ}, the Deutsch-Jozsa algorithm in qudit system is discussed and its unique extension, the Bernstein-Vazirani algorithm is provided in~\S\ref{subsubsec:quditBV}.

\subsubsection{Parity determining algorithm }
\label{sec:speedup alg}
 In this section we review a single qutrit algorithm which provides a two to one speedup than the classical counterpart. This algorithm can also be generalized to work on an arbitrary~$d$-dimensional qudit which solves the same problem of a larger computational space~\cite{Gedik2015}. In quantum computing, superposition, entanglement and discord are three important parts for the power of quantum algorithms and yet the full picture behind this power is not completely clear~\cite{Van_den_Nest2013}.
 
 Recent research shows that we can have a speedup in a fault tolerant quantum computation mode using the quantum contextuality ~\cite{Howard2014}. The contextual nature can be explained as ``a particular outcome of a measurement cannot reveal the pre-existing definite value of some underlying hidden variable''~\cite{Kochen1975,Klyachko2008}. In other words, the results of measurements can depend on how we made the measurement, or what combination of measurements we chose to do.
For the qudit algorithm discussed below, a contextual system without any quantum entanglement is shown to solve a problem faster than the classical methods~\cite{Gedik2015}.
 Because this qudit algorithm uses a single qudit throughout the process without utilizing any correlation of quantum or classical nature, it acts as a perfect example to study the sources of the quantum speed-up other than the quantum correlation. 

\begin{figure}[h!]
\begin{center}
\includegraphics[width=10cm]{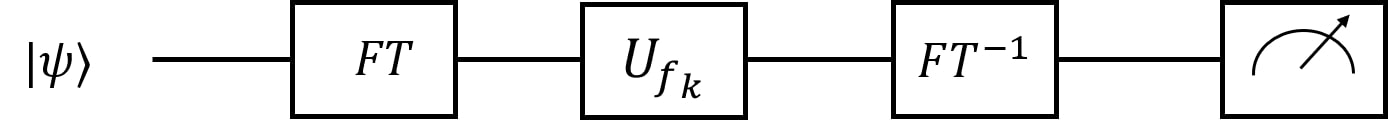}\\
\caption{Schematic view of the quantum circuit for the parity determining algorithm. $FT$ is the Fourier transform and~$U_{f_k}$ is the gate that does one of the two permutations and the last box represents the measurement. } 
\label{fig:qudit_effi}
\end{center}
\end{figure}

The algorithm solves a black-box problems that maps~$d$ inputs to~$d$ outputs after a permutation. Consider the case of three objects where six possible permutations can be divided into two groups: \textit{even} permutation that is a cyclic change of the elements and \textit{odd} permutation that is an interchange between two elements. If we define a function $f(x)$ that represents the permutation on the set $x\in \{-1,0,1\}$, the problems become  determining the parity of the bijection $f:{-1,0,1}\rightarrow{-1,0,1}$. We use Cauchy's two-line notation to define three possible even functions $f_k$, namely,
\begin{equation}
f_1:=\begin{pmatrix}
1&0&-1\\
1&0&-1
\end{pmatrix},\;
f_2:=\begin{pmatrix}
1&0&-1\\
0&-1&1
\end{pmatrix},\;
f_3:=\begin{pmatrix}
1&0&-1\\
-1&1&0
\end{pmatrix},
\end{equation}
and the remaining three odd function are
\begin{equation}
f_4:=\begin{pmatrix}
1&0&-1\\
-1&0&1
\end{pmatrix},\;
f_5:=\begin{pmatrix}
1&0&-1\\
0&1&-1
\end{pmatrix},\;
f_6:=\begin{pmatrix}
1&0&-1\\
1&-1&0
\end{pmatrix}.
\end{equation}
The circuit for the single qutrit algorithm in a space spanned by $\{\ket1,\ket0\,\ket{-1}\}$ is shown in Fig.~\ref{fig:qudit_effi}, where the operation~$U_{f_k}$ applies~$f_k$ to the state: $U_{f_k}(\ket1+\ket0+\ket{-1})=|f_k(1)\rangle+|f_k(0)\rangle+|f_k(-1)\rangle),$ and $FT$ is the single-qutrit Fourier transform
\begin{equation}
    FT=\frac{1}{\sqrt3}\begin{pmatrix}
\omega&1&\omega^{-1}\\
1&1&1\\
\omega^{-1}&1&\omega
\end{pmatrix}
\end{equation}
using~$\omega$ as the cube root of unity~(\ref{eq:omega}).
The process starts with state~$\ket1$ undergoing $FT$ and becoming~$\ket{\psi_1}$ as $FT\ket1=\ket{\psi_1}=\omega\ket1+\ket0+\omega^{-1}\ket{-1}$.
Then we obtain $\ket{\psi_k}$ by applying~$U_{f_k}$ to~$\ket{\psi_1}$. It is easy to show that
\begin{equation}
    \ket{\psi_1}
        =\omega^{-1}\ket{\psi_2}=\omega\ket{\psi_3}
\end{equation}
and, similarly,
\begin{equation}
    \ket{\psi_4}
    =\omega^{-1}\ket{\psi_5}=\omega\ket{\psi_6}.
\end{equation}
Hence, application of~$U_{f_k}$ on $\ket{\psi_1}$ gives $\ket{\psi_1}$ (up to a phase factor) for an even permutation and $\ket{\psi_4}=FT\ket{-1}$ for an odd permutation. Thus, applying inverse Fourier transform $FT^{-1}$ at the end, we measure $\ket1$ for even $f_k$ and $\ket{-1}$ for odd $f_k$. We are able to determine the parity of $f_k$ by a single application of $f_k$  on a single qutrit.

Generalizing to a~$d$–dimensional qudit system,
\begin{equation}
    \ket{\psi_k}
        :=\frac{1}{\sqrt{d}}\sum^d_{k'=1}
            \omega^{(k'-1)(k-1)}\ket{k'}.
\end{equation}
In this scenario,
a positive cyclic permutation maps $\ket{\psi_2}$ onto itself whereas negative  permutations give $\ket{\psi_d}$.
We then measure the results after applying an inverse Fourier transform to solve for the parity of the permutation. This algorithm has been implemented on the NMR system for both the qutrit~\cite{DOGRA2014} and ququart~\cite{Gedik2015} cases.
It is also realized on a linear optic system~\cite{Zhan2015}. Although the model problem has no significant applications and the speedup in the higher dimensional cases is not exponential, this proposed algorithm provides an elegant yet simple example for quantum computation without entanglement.

\subsubsection{Qudit Deutsch-Jozsa algorithm}
\label{subsubsec:quditDJ}
 Deutsch algorithm (with its origin in~\cite{Deu85} and improved in~\cite{CEMM98}) is one of the simplest examples to show the speed advantage of quantum computation. Deutsch-Jozsa algorithm is $n$-qubits generalization of the Deutsch algorithm. Deutsch-Jozsa algorithm can determine if a function $f(x)$ is \textit{constant}, with constant output, or \textit{balanced}, that gives equal instances of both outputs~\cite{Nielsen2011}. The process itself consists of only one evaluation of the function $f(x)$. In this algorithm,
 Alice sends Bob~$N$ qubits in the query register and one in the answer register where Bob applies the function to the query register qubits and stores the results in the answer register. Alice can measure the qubits in the query register to determine whether Bob's function is constant or balanced. This algorithm makes use of the superposition property of the qubit and reduces the minimum number of the function call from $2^n/2+1$ classically to only $1$ with quantum algorithm. This gives another example of the advantages of quantum algorithms.

The Deutsch-Jozsa algorithm can be performed in the qudit system with a similar setup.
Furthermore, with the qudit system, Deutsch-Jozsa algorithm can also find the closed expression of an affine function accurate to a constant term~\cite{Fan2007}.
The \textit{constant} and  \textit{balanced} function in the~$n$ dimensional qudit case have the following definition: ``An~$r$-qudit multi-valued function of the form
\begin{equation}
f:\{0,1,\ldots,n-1\}^r\to \{0,1,\ldots,n-1\}    
\end{equation}
is \textit{constant} when $f(x)=f(y)\; \forall x,y\in\{0,1,\ldots,n-1\}^r$ and is \textit{balanced} when an equal number of the $n^r$ domain values, namely $n^{r-1}$,
is mapped to each of the~$n$ elements in the co-domain''~\cite{Fan2007}.

It can be shown that all of the affine functions of~$r$ qudits
\begin{equation}
    f(x_1,\ldots,x_r):=A_0\oplus A_1x_1\oplus \cdots\oplus A_rx_r,\;
    A_0,\dots,A_r\in \mathbb{Z}_n,
\end{equation}
can be categorized to either constant or balanced functions~\cite{Fan2007}.
If all the coefficients $A_{i\neq 0}=0$ then the function is constant. For affine function with non-zero coefficient $A_{i\neq 0}$, every element in its domain $\{0,1,\ldots,n-1\}^r $ is reducible modulo~$n$ to a unique element $m \in\{0,1,\ldots ,n-1\}$. As $f(p)=f(q)$ if $p\equiv q (\text{mod}\,n)$, each of the elements in the codomain $\{0,1, \ldots , n-1\}$ is mapped to $n^{r-1}$ different elements in the domain. To finish the proof of the $n$-nary Deutsch-Jozsa algorithm, another trivial lemma is needed: Primitive $n^\text{th}$ roots of unity satisfy $\sum^{n-1}_{k=0}\omega^{\alpha k}=0$ for nonzero integers $\alpha$.

\begin{figure}[h!]
\begin{center}
\includegraphics[width=10cm]{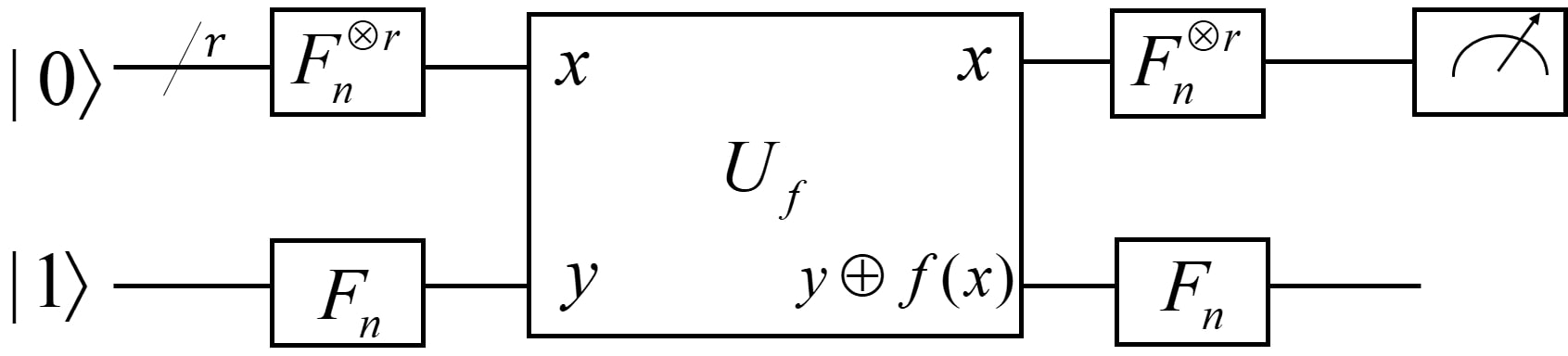}\\
\caption{The Deutsch-Jozsa circuit in qudit system. The $F_n$ are the qudit Hadamard gates achieved with quantum Fourier transform. } 
\label{fig:qudit_DeutschJozsa}
\end{center}
\end{figure}

The circuit of the Deutsch-Jozsa algorithm in qudits is shown in Fig.~\ref{fig:qudit_DeutschJozsa}. This algorithm of~$r$ qudits can both distinguish whether a function $U_f$ is balanced or constant and verify a closed expression for an affine function in $U_f$ within a constant term which is a universal phase factor of the $x$-register and thus is lost during the measurement.  The other coefficients of the affine function $A_1,\ldots,A_r$ are determined by measuring the state of the $x$-register at the output, $|A_1,\ldots,A_r\rangle$.

A detailed derivation of the circuit has been shown~\cite{Fan2007},
but the reasoning is an analogy to the qubit version of the Deutsch-Jozsa algorithm.
If the function $U_f$ is constant, the final state after the measurement is $\ket0^{\otimes r}\ket{n-1}$ as for $j\neq 0$ every states in the $x$-register have null amplitudes.
Therefore,
if every $x$-register qudit yields $\ket0$,
it is a constant function;
otherwise the function is balanced.

The Deutsch-Jozsa algorithm in the qudit system shares the same idea while enabling more applications such as determining the closed form of an affine function.
Although this algorithm is mainly of theoretical interest, the $n$-nary version of it may have applications in image processing.
It has the potential to distinguish between maps of texture in a Marquand chart since the images of which are encoded by affine functions~\cite{Nguyen2019}.
This algorithm can also be modified to set up a secure quantum key-distribution protocol~\cite{Nguyen2019}.
Other proposed Deutsch-Jozsa algorithms exist such as a method that makes use of the artificially allocated “subsystems” as qudits~\cite{KIKTENKO2015} and a generalized algorithm on the virtual spin representation~\cite{Kessel2002}.

\subsubsection{Qudit generalization of the Bernstein-Vazirani algorithm}
\label{subsubsec:quditBV}

in~\S\ref{subsubsec:quditDJ} we have discussed an application of a qudit Deutsch-Jozsa algorithm (DJA): verify a closed expression of an affine function.
This application is closely related to the Bernstein-Vazirani algorithm discussed in this section. Given an input string and a function that calculates the bit-wise inner-product of the input string with an unknown string, the Bernstein-Vazirani algorithm determines the unknown string~\cite{Bernstein1997}.
This algorithm can be treated as an extension of the Deutsch-Jozsa algorithm.

The qudit generalization of the Bernstein–Vazirani algorithm can determine a number string of integers modulo~$d$ encoded in the oracle function~\cite{KMS16,NGP+20}.
First we introduce a positive integer~$d$ and consider the problem in modulo~$d$ throughout. Given an $N$-component natural number string
\begin{equation}
    g(a):=(g(a_1),g(a_2),g(a_3),\ldots, g(a_N)),\;
    g(a_j)\in \{0,1,\ldots,d-1\},
\end{equation}
we define
\begin{equation}
    f(x):=g(a)\cdot x \mod d=g(a_1)x_1+g(a_2)x_2+\ldots +g(a_N)x_N \mod d,
\end{equation}
for
\begin{equation}
x=(x_1,x_2,\ldots,x_N)\in\{0,1,\ldots,d-1\}^N.    
\end{equation}
The oracle in the algorithm applies $f(x)$ to the input string $x$ and computes the result, 
namely, the number string $g(a)$ encoded in the function $f(x)$. 

The input state $x$ is chosen to be $\ket{\psi_0}=\ket0\otimes ^ N \ket{d-1}$, where $\ket0\otimes^N$ means initialization of the~$N$ control-qudits into their $\ket0$ states and $\ket{d-1}$ means the target qudit is in its $d-1$ state.
Quantum Fourier transforms of the pertinent input states are 
\begin{equation}
    \ket0\xrightarrow{QFT}\sum^{d-1}_{y=0}\frac{\ket{y}}{\sqrt{d}} 
\end{equation}
and 
\begin{equation*}
    \ket{d-1}\xrightarrow{QFT}\sum^{d-1}_{y=0}\frac{1}{\sqrt{d}}\omega^{d-y}\ket{y},
\end{equation*}
for~$\omega$ a root of unity~(\ref{eq:omega}).
The component-wise Fourier transform of a string encoded in the state~$\ket{x_1x_2\ldots x_N}$ is
\begin{equation}
   \ket{x_1x_2\ldots x_N} \xrightarrow{QFT}\sum_{z\in K} \frac{\omega^{x\cdot z}\ket{z}}{\sqrt{d^N}},
\end{equation}
where
\begin{equation}
    K=\{0,1,\ldots,d-1\}^N,\;
    z:=(z_1,z_2,\ldots,z_N).
\end{equation}

We denote the Fourier transform of the  $\ket{d-1}$ state as $\ket{\phi}$ and the input state after the Fourier transform is 
\begin{equation}
    \ket{\psi_1}=\sum_{x\in K} \frac{\ket{x}}{\sqrt{d^N}}\ket{\phi}
\end{equation}
Now we introduce the oracle as the $O_{f(x)}$ gate such that
\begin{equation}
    \ket{x}\ket{j}\xrightarrow{O_{f(x)}}\ket{x}\ket{(f(x)+j)\mod d},
\end{equation}
where 
\begin{equation}
    f(x)=g(a)\cdot x \mod d.
\end{equation}
By applying the $O_{f(x)}$ gate to $\ket{\psi_1}$ and following the formula by phase kick-back,
we obtain the output state
\begin{equation}
    O_{f(x)}\ket{\psi_1}
        =\ket{\psi_2}
        =\sum_{x\in K} \frac{\omega^{f(x)}\ket{x}}{\sqrt{d^N}}\ket{\phi}.
\end{equation}
Finally, obtain the $\ket{\psi_3}$ which is  the state after inverse Fourier transform of the first~$N$ qudits of $\ket{\psi_2}$. By measuring the first~$N$ quantum state of~$\ket{\psi_3}$  we can obtain the  natural number string we want that is offset up to a constant
\begin{equation}
    g(a_1),g(a_2),g(a_3),\ldots, g(a_N)
\end{equation}
 using a single query of the oracle function.

The Bernstein-Vazirani algorithm clearly demonstrates the power of quantum computing. It outperforms the best classical algorithm in terms of speed by a factor of~$N$~\cite{KMS16}. The qudit  generalizations of the Bernstein-Vazirani algorithm helps us comprehend the potential of the qudit systems.

\subsection{Qudit algorithms for the hidden Abelian subgroup problems. }
\label{sec:qudithiddenabelian}
 Many of the widely used quantum algorithms such as the discrete Fourier transform, the phase estimation and the factoring fit into the framework of the hidden subgroup problem (HSP). In this section, we review the qudit generalization of these algorithms. The qudit Fourier transform is discussed in~\S\ref{sec:qudit QFT} and its application, the PEA is reviewed in~\S\ref{sec:qudit PEA}. A direct application of these algorithms, Shor's factoring algorithm performed with qutrits and in metaplectic quantum architectures is also introduced ~\S\ref{sec:qudit PEA}.

\subsubsection{Quantum Fourier Transform with qudits}
\label{sec:qudit QFT} 
\begin{figure}[h!]
\begin{center}
\includegraphics[width=10cm]{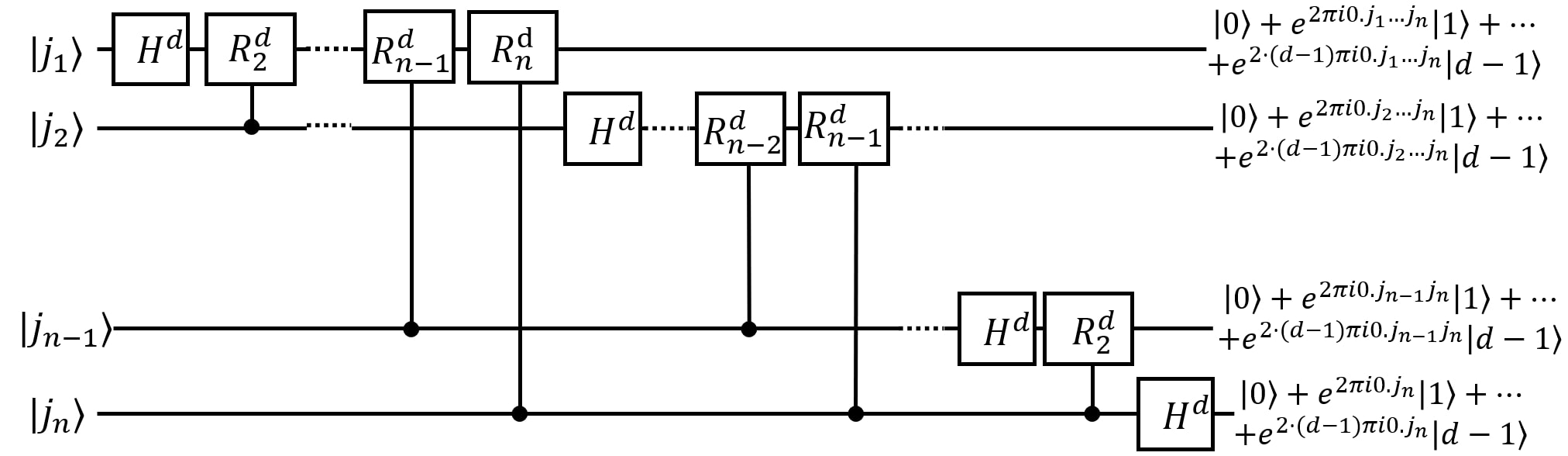}\\
\caption{Quantum Fourier transform in qudit system. $H^d$ is the~$d$-dimensional Hadamard gate and the expression of the $R^d$ gate is shown in Eq.~(\ref{eq:QFT RD}). 
Resultant states are shown to the right.  } 
\label{fig:qudit_QFT}
\end{center}
\end{figure}

The quantum Fourier transform algorithm (QFT) is realizable on a qubit system~\cite{Nielsen2011}. QFT, as the heart of many quantum algorithms, can also be performed in a qudit system~\cite{Muthukrishnan2002,Zilic2007}. In an $N$-dimensional system  represented with $n$~$d$-dimensional qudits, the QFT, $F(d,N)$, where $N=d^n$, transforms the computational basis \begin{equation}
    \{\ket0,\ket1,\ldots,\ket{n-1}\}
\end{equation}
into a new basis set~\cite{Cao2011}
\begin{equation}
F(d,N)\ket{j}=\frac{1}{\sqrt{N}}\sum^{N-1}_{k=0}\text{e}^{2\pi\text{i}j k /N}\ket{k}.
\end{equation}
For convenience, we write an integer~$j$ in a base-$d$ form. If $j>1$ then
\begin{equation}
    j=j_1j_2\cdots j_n=j_1d^{n-1}+j_2^{n-2}+\cdots+j_nd^0
\end{equation}
and, if $j<1$,
then
\begin{equation}
j=0.j_1j_2\cdots j_n
    =j_1d^{-1}+j_2d^{-2}+\cdots+j_nd^{-n}.    
\end{equation}
The QFT acting on a state~$\ket{j}$ can be derived and rewritten in a product form as
\begin{align*}
\ket{j}=\ket{j_1j_2\cdots j_n}&\mapsto \frac{1}{d^{n/2}}\sum^{d^n-1}_{k=0}\text{e}^{2\pi\text{i}j k /d^n}\ket{k}\\
&=\frac{1}{d^{n/2}}\sum^{d-1}_{k_1=0}\cdots\sum^{d-1}_{k_n=0}\text{e}^{2\pi\text{i}j(\sum^n_{l=1}k_ld^{-l})}|k_1k_2\cdots k_n\rangle\\
&=\frac{1}{d^{n/2}}\sum^{d-1}_{k_1=0}\cdots\sum^{d-1}_{k_n=0} \bigotimes^n_{l=1}\text{e}^{2\pi\text{i}j k_l d^{-l}}\ket{k_l}\\
&=\frac{1}{d^{n/2}}\bigotimes^n_{l=1}\left[ \sum^{d-1}_{k_l=0}\text{e}^{2\pi\text{i}j k_l d^{-l}}\ket{k_l}\right].
\end{align*}
This process can be realized with the quantum circuit shown in Fig.~\ref{fig:qudit_QFT}, and the fully expanded expression of the product form is shown on the right side of the figure. The generalized Hadamard gate $H^d$ in the figure is defined as $H^d:=F(d,d)$ which effects the transform
\begin{equation}
H^d|j_n\rangle=\ket0+\text{e}^{2\pi\text{i}0.j_n}\ket1+\cdots +\text{e}^{2(d-1)\pi\text{i}0.j_n}\ket{d-1}.
\end{equation}
The matrix representation of~$H^d$ is 
\begin{equation}
\begin{pmatrix}
1 & 1 & \cdots & 1\\
1 & \text{e}^{2\pi\text{i}0.1} & \cdots &\text{e}^{2\pi\text{i}0.(d-1)}\\
\vdots & \vdots & \ddots & \vdots\\
1 & \text{e}^{2(d-1)\pi\text{i}0.1}& \cdots &\text{e}^{2(d-1)\pi\text{i}0.(d-1)}
\end{pmatrix}.
\end{equation}
In the circuit the $R^d_k$ gate is a phase gate that has the expression
\begin{equation}
\label{eq:QFT RD}
R^d_k=
\begin{pmatrix}
1 & 0 & \cdots & 0\\
0 & \text{e}^{2\pi\text{i}/d^k} & \cdots &0\\
\vdots & \vdots & \ddots & \vdots\\
0 & 0& \cdots &\text{e}^{2\pi\text{i}(d-1)/d^k}
\end{pmatrix}.
\end{equation}

The black dots in the circuit are  multi-value-controlled gates that apply $R^d_k$ to the target qudit~$j$ times for a control qudit in state~$\ket{j}$.
In order to complete the Fourier transform and ensure the correct sequence of $j_1j_2\cdots j_n$, a series of SWAP gates are applied at the end, which are not explicitly drawn in Fig.~\ref{fig:qudit_QFT}.

The QFT developed in qudit system offers a crucial subroutine for many quantum algorithm using qudits.
Qudit QFT offers superior approximations  where the magnitude of the error  decreases exponentially with~$d$ and the smaller error bounds are smaller~\cite{Zilic2007},
which outperforms the binary case~\cite{Coppersmith1994}.

\subsubsection{Phase-estimation algorithm with qudits}
\label{sec:qudit PEA}
With the qudit quantum Fourier transform, we are able to generalize the PEA to qudit circuits~\cite{Cao2011}. Similar to the PEA using qubit, the PEA in the qudit system is composed by two registers of qudits. The first register contains $t$ qudits and $t$ depends on the accuracy we want for the estimation. We assume that we can perform a unitary operation~$U$ to an arbitrary number of times using qudit gates and generate its eigenvector~$\ket{u}$ 
and store it using the second register's qudits~\cite{BRS17}.  We want to calculate the eigenvalue of $\ket{u}$ where $U\ket{u}=\text{e}^{2\pi\text{i}r}\ket{u}$
by estimating the phase factor~$r$. 

\begin{figure}[h!]
\begin{center}
\includegraphics[width=15cm]{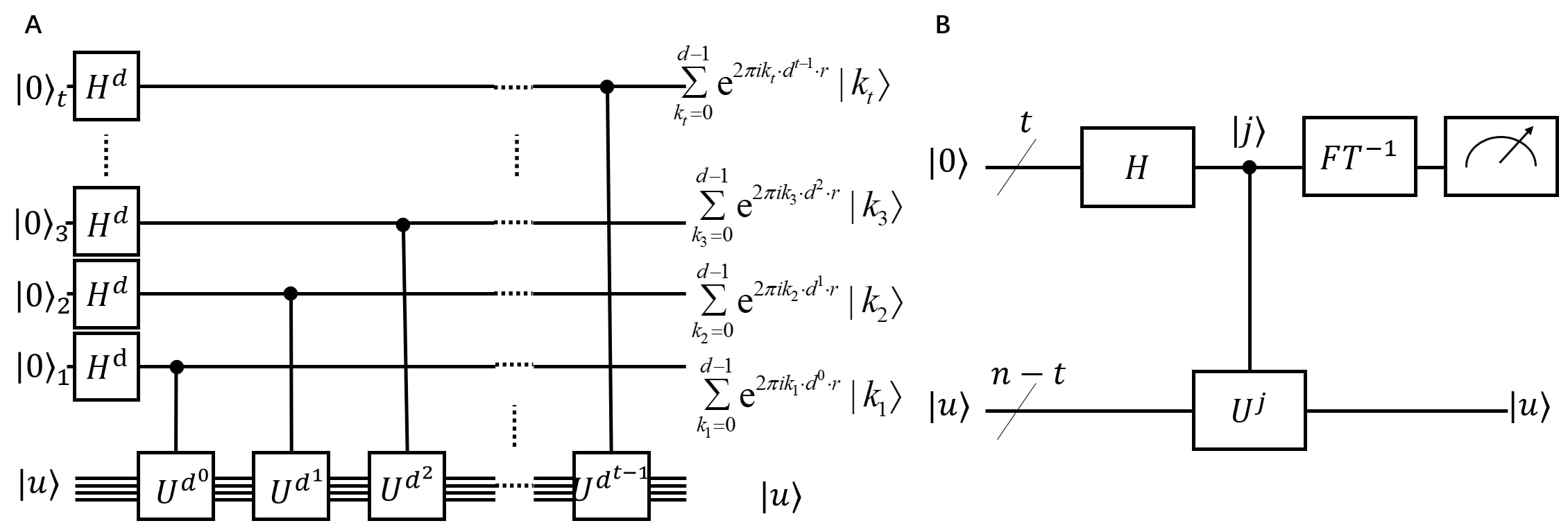}\\
\caption{\textbf{A} The circuit for the first stage of the PEA. The qudits in the second register whose states represent $\ket{u}$ are undergoing the~$U$ operations and the generated phase factors are kicking back to the qudits in the first register, giving the results to the right. \textbf{B} The schematic circuit for the whole stage of PEA. After the first stage of the PEA, inverse Fourier transform($FT^{-1}$) is applied to the qudits in the first register and the phase factors can be obtained by measuring the states of the first register qudits. } 
\label{fig:qudit_1_PEA}
\end{center}
\end{figure}

The following derivations follow those in Ref.~\cite{Cao2011}. For convenience, we rewrite the rational number~$r$ as 
\begin{equation}
    r=R/d^t=\sum^t_{l=1}\bar{R_l}/d^l=0.\bar{R_1}\bar{R_2}\cdots\bar{R_t}.
\end{equation}
As shown in Fig.~\ref{fig:qudit_1_PEA}~A, each qudit in the first register passes through the generalized Hadamard gate
$H\equiv F(d,d)$.
For the $l^\text{th}$ qudit of the first register, we have
\begin{equation}
\label{Fdd}
    F(d,d)\ket0_l=\frac{1}{\sqrt{d}}\sum^{d-1}_{k_l=0}\ket{k_l}.
\end{equation}
Then the $l^\text{th}$ qudit is used to control the operation $U^{d^{t-l}}$ on the target qudits of the state~$\ket{u}$ in the second register, which gives
\begin{equation}
    CU^{d^{l-1}}\ket{k}\otimes \ket{u}=\ket{k} (U^{d^{t-l}})^k\ket{u}=\text{e}^{2\pi\text{i}kd^{t-l}r} \ket{k}\otimes \ket{u}.
\end{equation}

Note that the function of the controlled operation $CU^{d^{t-l}}$ can be considered as a 'quantum multiplexer'~\cite{Bullock2005,Shende2006,KHAN2006336}. 
After executing all the controlled operations on the qudits, the qudit system state turns out to be 
\begin{equation}
\left(\prod^t_{l=1}\otimes\frac{1}{\sqrt{d}}\sum^{d-1}_{k_l=0} \text{e}^{2\pi\text{i}k_ld^{t-l}r} \ket{k_l}\right)\otimes \ket{u}.
\end{equation}
Therefore, through a process called the ``phase kick-back'',
the state of the first register receives the phase factor and becomes
\begin{equation}
    \ket{\text{Register 1}}=\frac{1}{d^{t/2}}\sum^{d^t-1}_{k=0}\text{e}^{2\pi\text{i}r k}\ket{k}.
\end{equation}
The eigenvalue~$r$ which is represented by the state $\ket{R}$ can be derived by applying the inverse QFT to the qudits in the first register:
\begin{equation}
F^{-1}(d,d^t)\ket{\text{Register 1}}
=\ket{R}.
\end{equation}
The whole process of PEA is shown in Fig.~\ref{fig:qudit_1_PEA}~ B. To obtain the phase $r=R/d^t$ exactly, we can measure the state of the first register in the computational basis.

The PEA in qudit system provides a significant improvement in the number of the required qudits and the error rate decreases exponentially as the qudit dimension increases~\cite{Parasa2011}.
A long list of PEA applications includes Shor's factorization algorithm~\cite{shor1994},
simulation of quantum systems~\cite{Lloyd1999},
solving linear equations~\cite{Lloyd2009,pan2014experimental},
and quantum counting~\cite{Tonchev2016}. To give some examples, a quantum simulator utilizing the PEA algorithm has been used to calculate the molecular ground-state energies~\cite{Aspuru-Guzik1704} and to obtain the energy spectra of molecular systems~\cite{Hefeng2008,daskin2011decomposition,daskin2012universal,kais2014introduction,bian2019quantum}. Recently, a method to solve the linear system using a qutrit version of the PEA has been proposed~\cite{sawerwain2013quantum}.
The qudit version of the PEA opens the possibility to realize all those applications that have the potential to out-perform their qubit counterparts.

Shor's quantum algorithm for prime factorization gives an important example of super-polynomial speed-up offered by a quantum algorithm over the currently-available classical algorithms for the same purpose~\cite{shor1999polynomial}.  The order-finding algorithm at the core of the factoring algorithm is a direct application of the PEA. With the previous discussion on the qudit versions of the quantum Fourier transform and phase estimation, we have the foundation to generalize Shor's factoring algorithm to the higher dimensional qudit system. 
Several proposals for performing Shor's algorithm on the qudit system, such as the adiabatic quantum algorithm of two qudits for factorization ~\cite{ZE12}, exist.
This method makes use of a time-dependent effective Hamiltonian in the form of a sequence of rotation operators that are selected accoding to the  qudit's transitions between its neighboring levels.

Another proposal carries out a  computational resource analysis on two quantum ternary platforms~\cite{BRS17}.
One is the ``generic" platform that uses magic state distillation for universality~\cite{CAB12}.
The other, known as a
metaplectic topological quantum computer (MTQC),
is a non-Abelian anyonic platform, where anyonic braiding and  interferomic measurement is used to achieved the universality with a relatively low cost~\cite{cui2015universal,cui2015universal2}.
The article discusses two different logical solutions for Shor's period-finding function on each of the two platforms: one that encodes the integers with the binary subspace of the ternary state space and optimizes the known binary arithmetic circuits; the other encodes the integer directly in the ternary space using the arithmetic circuits stemming in Ref~\cite{Bocharov2016ternary}.
Significant advantages for the MTQC platform are found compared to the others.
In particular the MTQC platform can factorize an $n$-bit number with $n+7$ logical qutrits with the price of a larger circuit-depth.
 To sum up the comparison, the  MTQC provides significant flexibility at the period finding algorithm for the ternary quantum computers.

\subsection{Quantum search algorithm with qudits}
\label{sec:qudti search}
The quantum search algorithm, also known as Grover's algorithm, is one of the most important quantum algorithms that illustrates the advantage of quantum computing. Grover's algorithm is able to outperform the classical search algorithm for a large database. The size of the computational space in an $n$-qubit system is a Hilbert space of $2^n$ dimensions.

Since there is a practical limit for the number of working qubits,
the working Hilbert space can be expanded by increasing the dimension of each carrier of information, i.e., using qudits and qudit gates. Several schemes of Grover’s quantum search with qudits have been proposed, such as one that uses the discrete Fourier transform as an alternative to the Hadamard gate~\cite{fan2008}
or another $d$-dimensional transformation~\cite{LI20114249} for the construction of the reflection-about-average operator (also known as the diffusion operator). In this section, an instruction on setting up Grover's algorithm in the qudit system is reviewed as well as a proposal of a new way to build a quantum gate $F$ that can generate an equal-weight superposition state from a single qudit state~\cite{Ivanov2012}. With the new gate $F$, it is easier to realize Grover's algorithm in a physical system and improve the overall efficiency of the circuit. 

\begin{figure}[h!]
\begin{center}
\includegraphics[width=15cm]{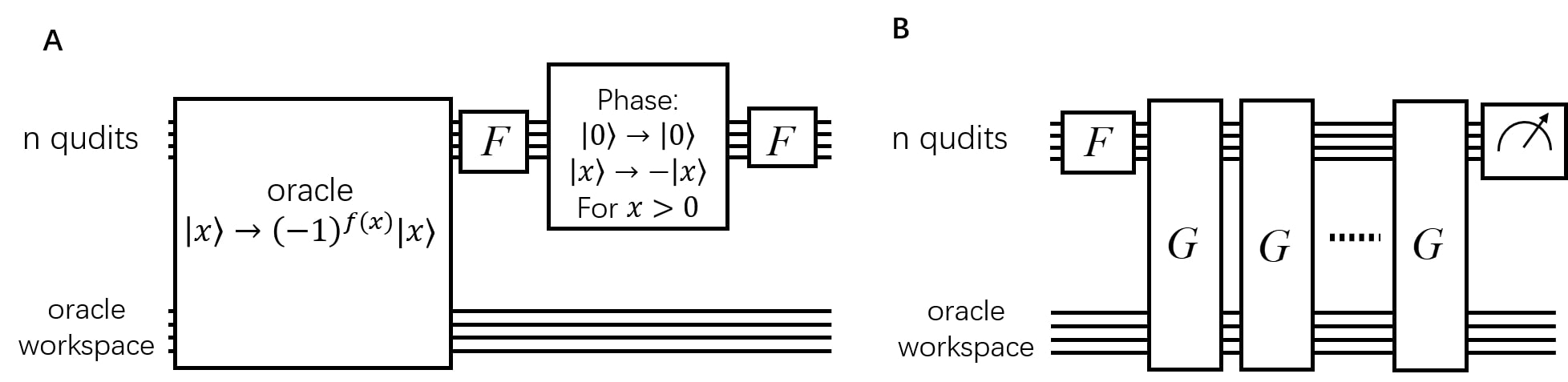}\\
\caption{\textbf{(A)} Circuit illustration for Grover iteration, $\bm{G}$, in a qudit system. The $F$ gate is the proposed qudit gate that transforms the single-qudit state $|0_k\rangle$ into an equal weight superposition state. \textbf{(B)} Schematic circuit illustration of the qudit quantum search algorithm. } 
\label{fig:qudit_grover}
\end{center}
\end{figure}

Grover's algorithm solves the unstructured search problem by applying Grover's oracle iteratively as shown in Fig.\ref{fig:qudit_grover}~B. To construct the oracle, we build qudit gates to perform the oracle function $f(x)$ that acts differently on the search target $s$ as compared to all the others.  The logic behind the algorithm is  to amplify the amplitude of the marked state~$\ket{s}$ with the oracle function, while attenuating the amplitudes of all the other states. The marked state is amplified enough to be located in $O(\sqrt{N})$ steps for an~$N$ dimensional search space. In each step Grover's oracle is executed one time. This oracle can be broken into two parts: (1)\textit{Oracle query.} The oracle shifts the phase of the marked state~$\ket{s}$ and leaving others unchanged by doing
\begin{equation}
\bm{R}_s(\phi_s)=\bm{1}+ (\text{e}^{\text{i}\phi_s}-1)\ket{s}\bra{s}.
\end{equation}
(2)\textit{Reflection-about-average.}  This operation is a reflection about a vector $\ket{a}$ with a phase $\phi_a$:
\begin{equation}
\bm{R}_a(\phi_a)=\mathbf{1}+ (\text{e}^{\text{i}\phi_a}-1)\ket{a}\bra{a}.
\end{equation}
It is constructed by applying the generalized Hadamard gate $H$, applying phase shift to $\ket0$ state and then applying $H$ again. It is straightforward to show that $$H^{\otimes n}\bm{R}_0(\phi_a)H^{\otimes n}=\bm{R}_a(\phi_a).$$ The two steps combined form Grover's operator $\bm{G}$, which is one execute of Grover's iteration. This process of Grover's iteration $\bm{G}$ is shown in Fig.\ref{fig:qudit_grover}~A.
 
\begin{figure}[h!]
\begin{center}
\includegraphics[width=10cm]{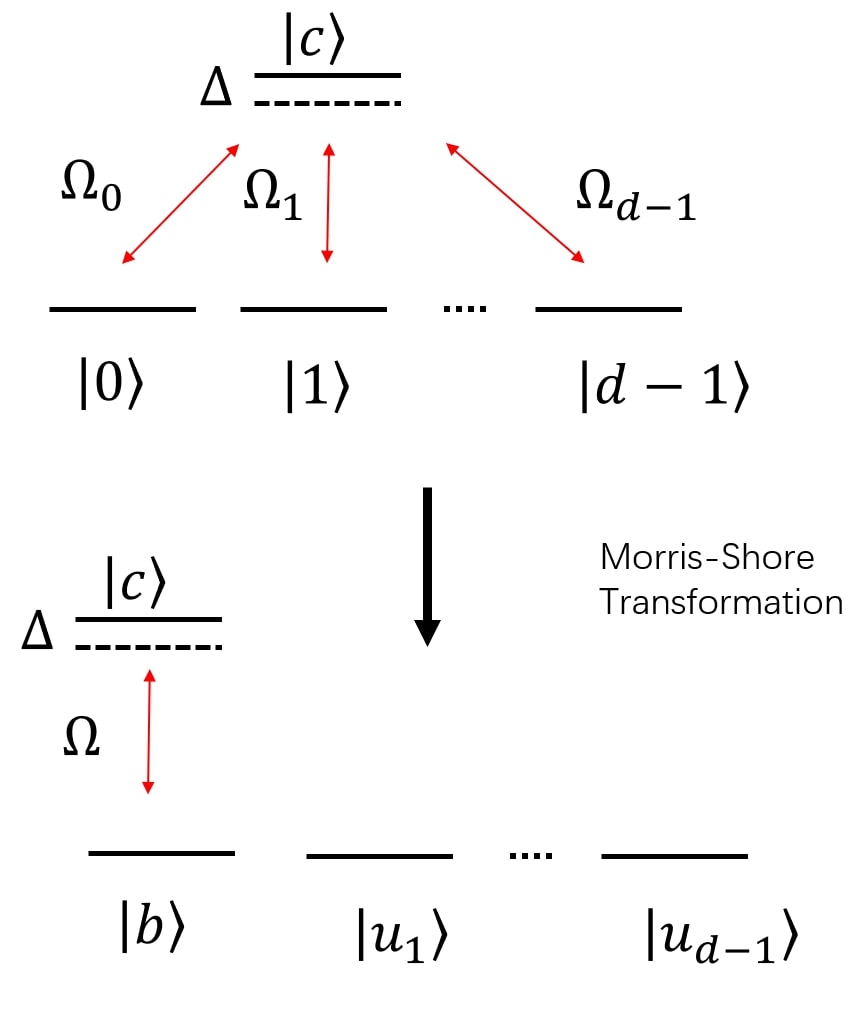}\\
\caption{Illustration of a qudit multipod linkage: the top is in the original basis and the bottom is in the Morris-Shore basis. $\Delta$ is a common detuning  between a common (ancilla)
state and other qudit states, $\Omega_k$ represents the single-photon Rabi
frequencies.  State $\ket{b}$ is a
superposition of the qudit states weighted by the couplings $\Omega_k$; $\ket{u_n}$
are  the states that are not in the dynamics. } 
\label{fig:qudit_search}
\end{center}
\end{figure}

Building Grover's operator in a qudit system can be simplified both algorithmically and physically. The most important improvement can be achieved by replacing the Hadamard gate $H$ with $F$ which drives the single-qudit state $|0_k\rangle$ into an equal weight superposition state,
\begin{equation}
F|0_k\rangle=\sum^{d-1}_{q=0}\xi_q|q_k\rangle,
\end{equation}
with $|\xi_q|=d^{-1/2}$, in all qudits $(k\in\{1,2,\ldots,n\}).$   The $F$ function can be realized by a single physical interaction in a multipod system easily. The multipod system  consists of $d$ degenerate quantum states
$\ket0,\ket1,\ldots,\ket{d-1}$. A common (ancilla) state $|c\rangle$ couples these states  to each other by two-photon
Raman processes, as illustrated in Fig.~\ref{fig:qudit_search}. The root-mean-square (rms) Rabi frequency as the coupling factor of the two states is
\begin{equation}
    \Omega(t)
    =\sqrt{\sum^{d-1}_{k=0}|\Omega_k(t)|^2}.
\end{equation}
Then from the two-state solution, we can calculate the dynamics of the multipod~\cite{Kyoseva2006}.

This method of building $F$ minimizes the number and the duration of algorithmic steps and thus is fast to implement and, in addition, it also provides better protection against detrimental effects such as decoherence or imperfections. Due to its conceptual simplicity, this method has applications in numerous physical systems. Thus, it is one of the most natural and simplest realizations of Grover's algorithm in qudits.

\section{Alternative models of quantum computing with qudits}
\label{subsec:alternative}
The gate-based description of quantum computing is useful to establish principles of quantum computing with qudits,
similar to the case for qubits.
There are various approaches to quantum computing besides the gate-based model,
such as the measurement-based~\cite{RB01},
adiabiatic quantum computing~\cite{FGGS00,AvDK+07}
and topological quantum computing~\cite{FLW02}.
Qudit versions of these approaches are barely explored to date, and we summarize the current status of these studies below.
\subsection{Measurement-based qudit computing}
\label{subsubsec:measurementbased}
Measurement-based quantum computing was introduced as an alternative approach to quantum computing whereby a highly entangled state,
such as a cluster state~\cite{BR01}
or its graph-state generalization~\cite{HEB04},
is prepared and then computation is performed by sequential single-qubit measurements in bases that are determined by a constant number of previous measurement outcomes~\cite{RB01,Nie06}.
Measurement-based quantum computing is appealing in settings where preparing a highly entangled many-qubit graph state is feasible,
such as parallelized controlled-phase operations~\cite{RB01}
or cooling to the ground-state of a special Hamiltonian~\cite{Nie06}.

Measurement-based qudit quantum computing is unexplored to date.
Preparatory work on generalizing graph states,
implicitly including the cluster-state special case,
to qudit graph states has been reported~\cite{KFMS10}.
Regarding implement,
qudit-based approaches have only been reported for the error-correction aspect of measurement-based qubit quantum computing~\cite{JLKK19}.
In this approach,
the cluster state is envisioned as comprising qudits,
with the high-dimensional nature of qudits serving to encode qubits for error correction.
They propose continuous-variable realizations
of a qudit cluster state in a continuous-variable setting~\cite{JLKK19}.
\subsection{Adiabatic qudit computing}
\label{subsubsec:adiabatic}
Adiabatic quantum computing approaches quantum computing by encoding the solution of a computational problem as the ground-state of a Hamiltonian whose description is readily obtained;
the solution is obtained by preparing the ground state of a Hamiltonian whose ground-state is efficiently constructed and then evolving slowly,
according to the adiabatic condition,
into a close approximation of the ground state of the Hamiltonian specifying the problem~\cite{FGGS00}.
The advantage of adiabatic quantum computing is evident in its natural correspondence to quantizing satisfiability problems~\cite{FGGS00},
and current efforts to exploit adiabatic quantum computing focus on quantum annealing,
which is a quantum generalization of the simulated annealing metaheuristic used for non-quantum global optimization problems~\cite{FGS+94,KN98,DC08}.

Quantum annealing is an important branch of quantum computing, particularly at the commercial level exemplified by D-Wave's early and continuing work in this domain.
As D-Wave researchers themselves point out,
realistic solid-state devices treated as qubits are not actually two-level systems and higher-dimensional representations of the dynamics must be considered to model and simulate realistic solid-state quantum annealers.
The effect of states outside the qubit space,
namely the treatment of solid-state quantum annealing as qudit dynamics,
has been studied carefully with conditions established for soundness of qubit approximations~\cite{ADS13}.

In fact the qudit nature of so-called superconduting qubits,
i.e., the higher-dimensional aspects of the objects serving as qubits,
is not just a negative feature manifesting as leakage error;
remarkable two-qubit gate performance is achieved by exploiting adiabatic evolution involving avoided crossings with higher levels~\cite{BKM+14,MG14}
with this exploitation for fast, high-fidelity quantum gates extendable to three-qubit gates and beyond by exploiting intermediate qudit dynamics and avoided level crossings~\cite{ZGS15,ZGS16}.
Another suggestion for exploiting qudit dynamics concerns using a degenerate two-level system
with the additional freedom perhaps improving the energy gap and thus increasing success probability~\cite{WSK20}.

A dearth of studies have taken place to date into qudit-based adiabatic quantum computing.
The one proposal thus far concerns a quantum adiabatic algorithm for factorization on two qudits~\cite{ZE12}.
Specifically,
they consider two qudits of possibly different dimensions,
thus necessitating a hybrid two-qudit gate~\cite{DWS03}.
They propose a time-dependent effective Hamiltonian 
to realize this two-qudit gate
and its realization as radio-frequency magnetic field pulses.
For this model,
they simulate factorization of each of the numbers 35, 21, and 15 for two quadrupole nuclei with spins 3/2 and~1, respectively, 
corresponding to qudit dimensions of 4 and~3, respectively.
\subsection{Topological quantum computing with qudits}
\label{subsubsec:topological}
Topological quantum computing offers advantages over other forms of quantum computing by reducing quantum error correction overheads by exploiting topological protection.
Some work has been done on topological quantum computing with qudits by proposing quantum computing with parafermions~\cite{HL16,DMCJ19}.

Majorana fermions are expected to exhibit non-abelian statistics,
which makes these exotic particles,
or their quasiparticle analogue,
sought after for anyonic quantum computing~\cite{Kit03}.
Majorana fermions can be generalized to~$\mathbb{Z}_d$ parafermions,
which also exhibit non-abelian statistics
and reduce to standard Majorana fermions for $d=2$.
One advantage of $d>2$ is that parafermion braiding is an entangling operation.
Importantly,
encoding a qudit of dimension~$d$
in the four-parafermion fusion space enables all single-qudit Clifford gates to be generated modulo phase terms~\cite{HL16}.

Clifford gates do not provide a universal set of gates for quantum computing.
A non-Clifford gate can be achieved for parafermions encoded into parafermion zero modes by exploiting the Aharonov-Casher effect,
physically implemented by move a half-fluxon around the parafermionic zero modes.
Combining this non-Clifford gate with the Clifford gates achieved by parafermion braiding yields a universal gate set of non-abelian quantum computing with qudits~\cite{DMCJ19}.

\section{Implementations of qudits and algorithms}
\label{sec:implementationsqudits}

The qubit circuit and qubit algorithm have been implemented on various physical systems such as defects in solids~\cite{Jelezko2004,Childress2006,Neumann2010}, quantum dots~\cite{Loss1998,Nowack2007}, photons~\cite{Milburn_2009,Prevedel2007}, super conducting systems~\cite{Chiorescu2003,Clarke2008}, trapped ions~\cite{Blatt2008,Bloch2008}, magnetic~\cite{troiani2011,bogani2010,clemente2012magnetic,aromi2012design} and non-magnetic molecules~\cite{chuang1998experimental,vandersypen2001experimental}.
For each physical representation of the qubit, only two levels of states are used to store and process quantum information. However, many quantum properties of these physical systems have more than two levels, such as the frequency of the photon~\cite{Hsuan-Hao2019}, energy levels of the trapped ions~\cite{klimov2003}, spin states of the nuclear magnetic resonance systems~\cite{DOGRA2014} and the spin state of the molecular magnetic magnets~\cite{Moreno-Pineda2018}. Therefore, these systems have the potential to represent  qudit systems. In this section, we briefly review several physical platforms that have been used to implement qudit gates or qudit algorithms.

Although most of the systems have three or four levels available for computation, they are extensible to higher level systems and scalable to multi-qudit interactions. These pioneer implementations of qudit systems show the potential of  future realization of the more powerful qudit quantum computers that have real-life applications. 
\subsection{Time and frequency bin of a photon}
\label{subsec:timefreqphoton}
Photonic system is a good candidate for quantum computing because photons rarely interact with other particles and thus have a comparatively long decoherence time. In addition, photon has many quantum properties such as the orbital angular momentum~\cite{Babazadeh2017,EFKZ18}, frequency-bin~\cite{KRR+17,lu2018electro,IJO+18,IJA+19} and time-bin~\cite{islam2017,Humphreys2013}  that can be used to represent a qudit. Each of these properties provides an extra degrees of freedom for the manipulation and computation.  Each degree of freedom usually has dimensions greater than two and thus can be used as a unique qudit. The  experimental realization of arbitrary multidimensional multiphotonic transformations has been proposed with the help of ancilla
state, which is achievable via the introduction of a new quantum nondemolition measurement and the exploitation of a genuine high-dimensional interferometer~\cite{GEZK19}.
Experimental entanglement of high-dimensional qudits, where  multiple high-purity frequency modes of the photons are in a superposition coherently, is also developed and demonstrated~\cite{KRR+17}.

Here we review a single photon system that has demonstrated a proof-of-principle qutrit PEA~\cite{Hsuan-Hao2019}. In a photonic system, there is no deterministic way to interact two photons and thus it is hard to build a reliable controlled gate for the photonic qudits. The following photonic system bypasses this difficulty via using the two degrees of freedom on a single photon---i.e., the time-bin and frequency-bin to be the two qutrits. The frequency degree of freedom carries one qutrit as the control register and the time degree of freedom carries another qutrit as the target register. The experimental apparatus consists of the well-established techniques and fiber-optic components: continuous-wave(CW) laser source, phase modulator(PM), pulse shaper(PS), intensity modulator(IM)  and chirped fiber Bragg grating(CFBG).
The device is divided into three parts~\cite{Hsuan-Hao2019}:
1.~A state preparation part that comprises a PM followed by a PS and a IM that encodes the initial state to qudits;
2.~a controlled-gate part that is built with a PM sandwiched by two CFBGs to perform the control-$U$ operation; and
3.~an inverse Fourier transformation comprising a PM and then a PS to extract the phase information.
Note that the controlled-gate part can perform a multi-value-controlled gate that applies different operations based on the three unique states of the control qutrit.
In the PEA procedure,
eigenphases can be retrieved with $98\%$ fidelity.
In addition to having long coherence lifetime, the photonic system also has a unique advantage over other common quantum devices  i.e., the ability to process and measure thousands of photons simultaneously. This allows us to generate statistical patterns quickly and infer the phase accurately whereas the normal PEA has to use additional qudits on the control register to increase accuracy.

\begin{table}[ht]
    \centering
    \begin{tabular}{|c||c|c|c|}
    \hline
    \multicolumn{4}{|c|}{$\hat{U}_1$}   \\ \hline
    \textbf{Eigenstate}             
   &$\boldsymbol{\ket0_t}$       &$\boldsymbol{\ket1_t}$         &$\boldsymbol{\ket{2}_t}$       \\ \hline \hline 
    $\boldsymbol{E_0}$                           
   &$.9948\pm.0004$   &$.0101\pm.0004$     &$.0122\pm.0005$\\ \hline
    $\boldsymbol{E_1}$                           
   &$.0023\pm.0002$    & $.9805\pm.0009$      & $.0120\pm.0005$\\ \hline
    $\boldsymbol{E_2}$                           
    & $.0029\pm .0002$   & $.0094\pm.0004$      & $.9758\pm.0010$\\ \hline
    \textbf{True Phase}, $\phi$     
    & $0$& $2\pi/3$           & $4\pi/3$        \\ \hline
    \textbf{Est. Phase}, $\tilde{\phi}$    
    & $1.972\pi$        & $.612\pi$           & $1.394\pi$       \\ \hline
    \textbf{Error}, $\frac{|\phi - \tilde{\phi}|}{2\pi}$  
    & $1.4\%$           & $2.7\%$             & $3.0\%$             \\ \hline
    \hline \hline
    \multicolumn{4}{|c|}{$\hat{U}_2$}   \\ \hline
    \textbf{Eigenstate}             
    & $\boldsymbol{\ket0_t}$        & $\boldsymbol{\ket1_t}$          & $\boldsymbol{\ket{2}_t}$       \\ \hline \hline
    $\boldsymbol{E_0}$                           
    & $.878\pm.002$      & $.316\pm.003$        & $.143\pm.002$\\ \hline
    $\boldsymbol{E_1}$                           
    & $.032\pm.001$      & $.530\pm.003$        & $.318\pm.003$\\ \hline
    $\boldsymbol{E_2}$                           
    & $.090\pm .002$     & $.154\pm.002$        & $.539\pm.003$\\ \hline
    \textbf{True Phase}, $\phi$     
    & $0$                & $.3511\pi$           & $1.045\pi$        \\ \hline
    \textbf{Est. Phase}, $\tilde{\phi}$   
    & $1.859\pi$        & $.377\pi$           & $1.045\pi$       \\ \hline
    \textbf{Error}, $\frac{|\phi - \tilde{\phi}|}{2\pi}$  
    & $7.1\%$           & $1.3\%$             & $0.0\%$             \\ \hline
    \end{tabular}
    \caption{Normalized photon counts and comparison of the true phase $\phi$ and the experimentally estimated phase $\phi'$  for each eigenstate of $\hat{U}_1$ (Eq.~\ref{unitary}) and $\hat{U}_2$ (Eq.~\ref{Unitary2})~\cite{Hsuan-Hao2019}. }
    \label{tab:Results}
\end{table}

Here we provide an example for the statistical inference of the phase based on numerical data generated by the photonic PEA experiment just described.
The two unitary operations used in the experimental setup are
\begin{equation}
    \hat{U}_1=\text{diag}(1,\omega,\omega^2),
\label{unitary}
\end{equation}
with~$\omega$ being the cube root of unity~(\ref{eq:omega}),
and
\begin{equation}
    \hat{U_2}
        =\text{diag}\left(1,\text{e}^{\text{i}{0.351\pi}},\text{e}^{\text{i}{1.045\pi}}\right).
\label{Unitary2}
\end{equation}
In the experiment, photonic qutrits are sent through the control and target registers and the state of the control register qutrits is measured and counted to obtain the phase information.

Given the eigenphase~$\phi$  of an eigenstate of the target register, the probability for the qutrit output state to fell into $\ket{n}$, where $n\in\{0,1,2\}$, is
\begin{align}
    C(n,\phi)=\frac{1}{9}\left|1+\text{e}^{\text{i}(\phi -\frac{n2\pi}{3})}+\text{e}^{\text{i}2(\phi -\frac{n2\pi}{3})}\right|^2.
\label{Eq:controlProb}
\end{align}
Now let $E_0$, $E_1$, and $E_2$ be the counts of the photons that fell into $\ket0_f$, $\ket1_f$, and $\ket{2}_f$. The estimated phase, denoted $\tilde{\phi}$, is the phase that has the smallest the mean-square error between the measured and theoretical results:
\begin{equation} \label{Eq:min}
    \min_{\tilde{\phi}} \sum_{n=0}^2 (E_n - C(n,\tilde{\phi}))^2
\end{equation}
The estimated phases for $\hat{U}_1$~(\ref{unitary}) and $\hat{U}_2$~(\ref{Unitary2}) are shown in Table~\ref{tab:Results}~\cite{Hsuan-Hao2019}. The first experiment with $U_1$ estimates the phase of a eigenvector and gives the eigenvalue. The second experiment with $U_2$ estimates the phase of a state with an arbitrary value (not a fraction of $\pi$ ), but,
by repeating the experiment,
the eigenvalue can be estimated from the statistical distribution of the results.

\subsection{Ion trap}
\label{subsec:iontrap}
Intrinsic spin, an exclusively quantum property, has an inherently finite discrete state space which is a perfect choice for representing qubit or qudit.
When a charged particle has spin,
it possess a magnetic momentum and is controllable by external electromagneic pulses.
This concept leads to the idea of ion trap where a set of charged ions are confined by electromagnetic field. The hyperfine (nuclear spin) state of an atom, and lowest level vibrational modes (phonons) of the trapped atoms serves as good representations of the qudits.
The individual state of an atom is manipulated with laser pulse and the ions interact with each other via a shared phonon state. 

\begin{figure}
\begin{center}
\includegraphics[width=10cm]{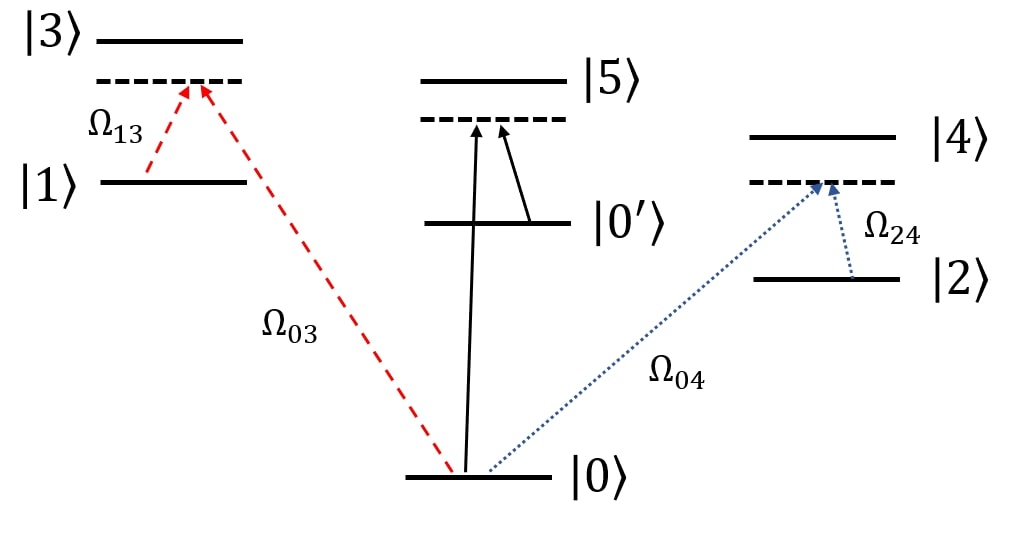}\\
\caption{Electronic level structure of the trapped ion. The carrier of the quantum information
is the qutrit states $\ket0$, $\ket1$, and $\ket2$. $\ket{0^{\prime}}$ is an auxiliary level used for the
conditional two-qutrit gate.} 
\label{fig:qudit_ion}
\end{center}
\end{figure}

The set-up of an ion trap qutrit system reviewed here can perform arbitrary single qutrit gates and a control-not gate~\cite{klimov2003}. These two kinds of gates form a universal set and thus can be combined to perform various quantum algorithms such as those discussed in~\S\ref{sec:qudit alg}. The electronic levels of an ion are shown in Fig.~\ref{fig:qudit_ion}. The energy levels $\ket0,\ket1,\ket2$ are used to store the quantum information of a qutrit. The transition between the levels are driven by the classical fields $\Omega_{03},\Omega_{13},\Omega_{04}$ and $\Omega_{24}$ of the Raman transitions through independent channels linked to orthogonal polarizations. We first develop a system acting as a single qutrit gate that can manipulate the energy levels of the ion via Raman transitions driven by the classical fields. 
The following expressions follow those in Ref.~\cite{klimov2003}.
For single qutrit gates, where the center-of-mass motion is excluded, we can include the spatial dependence of the Raman fields as phase factors $\Delta$ and assuming the conditions
\begin{equation}
    \Delta\gg\Omega_{04},\Omega_{03},\Omega_{31},\Omega_{42},
\end{equation}
the effective Hamiltonian describing the ion in this system  is
\begin{align}
    \frac{H}{\hbar}
    =&-\frac{|\Omega_{31}|^2}{\Delta}\ket1\bra1-\frac{|\Omega_{42}|^2}{\Delta}\ket2\langle 2|-\frac{|\Omega_{30}|^2+|\Omega_{40}|^2}{\Delta}\ket0\bra0-\\
&-\left[\frac{\Omega_{31}\Omega_{30}^{\ast}}{\Delta}\ket0\bra1+\frac{\Omega_{42}\Omega_{40}^{\ast}}{\Delta}\ket0\langle 2|+\text{hc} \right].
\end{align}
Knowing the Hamiltonian we are able to derive the evolution operator in the restricted three-dimensional space spanned by $\{\ket2,\ket1,\ket0\}$  as the following 
\begin{equation}
U(\varphi)=\begin{pmatrix}
1+|g|^2C(\varphi)&gg^{\prime\ast}C(\varphi)&-ig\sin\varphi  \\
g'g^{\ast}C(\varphi)& 1+|g'|^2C(\varphi)&-ig'\sin\varphi\\
-\text{i}g^{\ast}\sin\varphi&-\text{i}g^{\prime\ast}\sin\varphi&\cos\varphi
\end{pmatrix},
\end{equation}
where $\varphi=\Omega t$ represents interaction time
and 
\begin{equation}
    C(\varphi)=\cos \varphi-1,\;
    \Omega^2=|\kappa'|^2+|\kappa|^2.
\end{equation}
The notation $g$ and $g'$ represents
\begin{equation}
    g:=\kappa/\Omega,\;
    g'=\kappa'/\Omega,\;
    \kappa:=\Omega_{42}^{\ast}\Omega_{40}/\Delta,\;
    \kappa'=\Omega_{31}^{\ast}\Omega_{30}/\Delta.
\end{equation}
This evolution operator can perform all kinds of the required coherent operations that are acting on any two of the logical states. It operates on the system and works essentially as a single qutrit gate.
All kinds of transitions can be realized by manipulating the $\kappa$ and $\kappa'$ coupling. Therefore with the proper manipulation of the parameters $\kappa$ and $\kappa'$ we are able to perform any arbitrary one-qutrit gate as desired.

Single qutrit gate alone is not sufficient to form a universal computational set, as we need a conditional two-qutrit gate or a two-qutirt controlled-gate to achieve universality.
To define the conditional two-qutrit gate we need an auxiliary level $|0'\rangle$ as shown in Fig.~\ref{fig:qudit_ion}.  
The conditional two-qutrit gate is achievable via the center-of-mass (CM) motion of ions inside the trap.
The ion CM coupled to the electronic transition $\ket0\rightarrow\ket{q}$ is described by the Hamiltonian
\begin{equation}
    H_{n,q}=\frac{\Omega_q \eta}{2}[\ket{q}_n\bra0a \text{e}^{-\text{i}\delta t -i\phi}+a^\dagger\ket0_n\langle q|\text{e}^{\text{i}\delta t+\text{i}\phi}].
\end{equation}
Here $a$ is the annihilation operator and $a^\dagger$ is the creation operator of the CM phonons.  $\Omega_q$ is the effective Rabi frequency after adiabatic elimination of upper excited levels and $\phi$ is the laser phase, and $\delta$ is the detuning.
The Lamb-Dicke parameter is
\begin{equation}
    \eta
        :=\sqrt{\hbar k^2_{\theta}/(2M\nu_x)}.
\end{equation}
This Hamiltonian governs the coherent interaction
between qutrits and collective CM motion. 
With appropriate selection of effective interaction time and laser polarizations,
the CM motion coupled to electronic transitions is coherently manipulated~\cite{klimov2003}.

To complete the universal quantum computation requirements, we need to develop a measurement scheme.
In this scheme, von~Neumann measurements distinguishing three directions $\ket0$,  $\ket1$, $\ket2$ are made possible via the resonant interactions from $\ket1$ and $\ket2$ to states $\ket3$ and $\ket4$, respectively. The single and two-qutrit controlled gate  are combined to perform various qutrit algorithms such as the quantum Fourier transform. 
Other variations of the ion-trap qutrit quantum computer designs use trapped ions in the presence of a magnetic field gradient~\cite{mc2005trapped}. 
The qutrit ion-trap computer provides a significant increase of the available Hilbert space while demanding only the same amount of physical resources. 

\subsection{Nuclear magnetic resonance}
\label{subsec:nmr}
Nuclear magnetic resonance (NMR) is an essential tool in chemistry 
and involves manipulating and detecting molecules' nuclear spin states using radio-frequency electromagnetic waves~\cite{bovey1988nuclear}.
Some technologies of this field are sophisticated enough to control and observe thousands of nuclei in an experiment.
The NMR has the potential to scale up quantum computer to thousands of qudits~\cite{slichter2013principles}.

In this section we review the implementation of a single-qudit algorithm that can determine the parity of a permutation on an NMR system~\cite{DOGRA2014}.
The algorithm itself is the parity determining algorithm explained in ~\S\ref{sec:speedup alg}.
The molecule in this NMR setup is embedded in a liquid crystalline environment and the strong magnetic field is used to adjust the anisotropic molecular orientation. This addind  a finite quadrupolar coupling term to the Hamiltonian which is as follows
\begin{equation}
    H=-\omega_0I_z+\Lambda(3I_z^2-I^2),
\end{equation}
where $\Lambda=e^2qQS/4$ is the effective value of the quadrupolar coupling~\cite{DOGRA2014}.
The Fourier transformation is implemented by a sequence of three transition-selective pulses.
A series of  combinations of $180^{\circ}$ pulses, both transition-selective  and non-selective, is used to implement the permutations.

Final states of the system can be derived from a single projective measurement. Pseudopure spin states act as approximation of effect of the system  on an ensemble NMR quantum computer since it is impossible to do the true projective measurements~\cite{lee2006projective}.
The fidelity measurement of the experiment is given as
\begin{equation}
    F:=\frac{\text{tr}(\rho^\dagger_\text{th}\rho_\text{expt})}{\sqrt{\text{tr}(\rho^\dagger_\text{th}\rho_\text{th})}\sqrt{\text{tr}(\rho^\dagger_\text{expt}\rho_\text{expt})}}
\end{equation}
is used,
where~$\rho_\text{th}$ and~$\rho_\text{expt}$ are, respectively, theoretically expected and experimentally obtained density matrices.
Fidelities obtained for these proposed operations are 0.92 and above.

Another set-up of the same algorithm treats a single ququart~\cite{Gedik2015}.
The algorithm implementation is achieved using a spin–$\frac32$ nuclei, which is commonly selected for NMR-QIP applications. In their NMR systems the four energy levels needed is made via the Zeeman splitting using a strong static magnetic field. All of the two implementations of the single-qudit algorithm show that the NMR system provides a way to realize a reliable and efficient qudit system for the quantum computing.

\subsection{Molecular magnets}
\label{subsec:moluclarmagnets}
Molecular quantum magnets, also called the single-molecule magnets (SMM), provides another physical representation of  qudits~\cite{Moreno-Pineda2018}. They have phenomenal  magnetic characteristics and can be manipulated via chemical means. This enables the alternation of the ligand field of the spin carriers and the interaction between the SMM with the other units.
As pointed out in one of the proposals, the nuclear spin states of the molecules, which have a long life-time, are used to store the quantum information. This information is read out by the electronic states. In the mean time, the robustness of the molecule allows it to conserve its molecular, electronic and magnetic characteristics at high temperatures~\cite{Moreno2016}.

As one of the SMMs, the single molecule $\mathrm{TbPc_2}$ complex reviewed in this section possesses all necessary properties such as long lifetime and robustness.
These properties are integrated as important components of a serious quantum mechanical devices, for examples, resonator~\cite{ganzhorn2013strong}, molecular spin valve~\cite{urdampilleta2011} and transistor~\cite{sawerwain2013quantum,thiele2014}. 
 $\mathrm{TbPc_2}$ gains its SMM properties from the strong spin–orbit coupling of lanthanide ions and the ligand field~\cite{ishikawa2003}.
Magnetic properties of $\mathrm{TbPc_2}$ are governed by the Hamiltonian:
\begin{equation}
    \mathscr{H}
    =\mathscr{H}_{\mathrm{lf}}+g_J\mu_0\mu_B J\cdot H+A_{\mathrm{hf}}I\cdot J +(I^2_z-\frac13(I+1)I),
\end{equation}
where $\mathscr{H}_{\mathrm{lf}}$ is the ligand field Hamiltonian(lf),
and $g_J\mu_0\mu_B J\cdot H$ represents the Zeeman energy.
$A_{\mathrm{hf}}I\cdot J $ accounts for hyperfine interactions(hf) and $(I^2_z-\frac13(I+1)I) $ is the quadrupole term.
A sweeping magnetic field associated with $m_I=\pm\nicefrac12$ and $\pm\nicefrac32$ can cause quantum tunnelling of magnetisation, which preserves  nuclear spin while changing electronic magnetic moment. 
This field enables nuclear-spin measurement by suspending the $\mathrm{TbPc_2}$ molecule on carbon nanotubes (CNT) and between gold junctions.

This measurement uses the technique of electro-migration.
Initialisation and manipulation of the four spin states of $\mathrm{TbPc_2}$ can be obtained from QTM transitions driven by external ramping magnetic field.
The transitions between the $\ket{+\nicefrac12}\leftrightarrow\ket{-\nicefrac12}$ states and $\ket{+\nicefrac32}\leftrightarrow\ket{-\nicefrac32}$
is achieved via applying appropriate resonate frequencies $\nu_{12}$ and $\nu_{23}$.
Relaxation and coherence times are important aspects to be analyzed for the $\mathrm{TbPc_2}$ system,
and this process is accomplished by  imaging  the
initialized nuclear spin trajectory in real-time.

Statistical analysis of the nuclear spin coherence time makes use of the spin–lattice relaxation times by fitting the data for an exponential form $(y=\exp(-t/T_1))$ and yields $T_1\approx 17$~s for $m_I=\pm\nicefrac12$ and $T_1\approx 34s$ for $m_I=\pm\nicefrac32$ with fidelities of $F(m_I=\pm\nicefrac12)\approx93\%$ and $F(m_I=\pm\nicefrac32)\approx87\%$ accordingly~\cite{Moreno-Pineda2018}.
The $\mathrm{TbPc_2}$ SMM
can be used to execute Grover's algorithm, where the alternation of the $m_I$ state contained in the $\mathrm{TbPc_2}$ molecular qubit are treated by resonance frequencies~\cite{leuenberger2001,godfrin2017}.

\section{Summary and future outlook of qudit system}
\label{SEC:qudit sum}

\subsection{Summary of the advantages of qudit systems compared to qubit systems}

Throughout the article we discuss and review many aspects of the qudit systems such as qudit gates, qudit algorithms, alternative computation models and implementations. Most gates and algorithms based on qudits have some advantages over those for qubits,
such as shorter computational time, lower requirement of resources, higher availability, and the ability to solve more complex problems. 
The qudit system, with its high-dimensional nature, can provide more degrees of freedom and larger computational space. This section summarizes the advantages of the qudit system compared to the qubit system.

Qudit gates have the advantage of a larger working Hilbert space which reduces the number of qudits needed to represent an arbitrary unitary matrix. In our discussion of universality in \S\ref{subsubsec:universality_eg}, the qudit method proposed by  Muthukrishnan and Stroud's has a $(\log_2d)^2$ scaling advantage over the qubit case. Furthermore, Luo and Wang show that with their proposed universal computation scheme~\cite{Luo2014}, there is an extra factor of $n$ reduction in the gate requirement, where  $n$ is the number of qudits. By introducing qudits to the construction of some well-known gates such as the Toffoli gate, the elementary gate required are reduced from $12n-11$ gates in the qubit case to $2n-1$ gates by introducing a single ($n+1$)-level target carrier~\cite{Ralph2007} and to $2n-3$ gates by utilizing the topological properties~\cite{KNX+20}. In our discussion of the geometrically quantified qudit-gate efficiency in \S\ref{sec:gategeometry}, the qubit system needs $O(n^6d(I, U)^3)$ one- and two-qubit gates to synthesize a unitary~\cite{nielsen2005geometric} while in the qutrit case the lower bound is $O(n^kd(I, U)^3)$ where $k$ is an integer that depends on the accuracy of the approximation and can be smaller than $6$~\cite{li2013qutrits}. 

For many of the physical systems such as  photons~\cite{Milburn_2009,Prevedel2007}, super conducting systems~\cite{Chiorescu2003,Clarke2008}, trapped ions~\cite{Blatt2008,Bloch2008}, magnetic~\cite{troiani2011,bogani2010,clemente2012magnetic,aromi2012design} and non-magnetic molecules~\cite{chuang1998experimental,vandersypen2001experimental} there are usually more than two available physical states available for the applications. The qudit system has a higher efficiency utilizing those extra states than the qubit system. Also using the photonic system, we can perform the multi-level controlled gate (\S\ref{sec:qudit_MVCG}) which can perform multiple control operations at and same time and largely reduce the number of controlled gates requirement~\cite{Hsuan-Hao2019}. 

Other than computation, the qudit also has advantages in quantum communication as it possesses a higher noise resilience than the qubit~\cite{cozzolino2019high}. The qudit system has a higher quantum bit error rate (QBER), which is a measure of resistance to the environmental noise or eavesdropping attacks, compared to the qubit system. The higher noise tolerance of the qudits helps to increase the secret key rate as it can be shown that the secret key rate increases
as the Hilbert space dimensions increase at the same noise level~\cite{sheridan2010security}. Notice that in practical situation, the qudit system performed on each particular physical apparatus has varied amount of advantages than the qubit and there might be cases in which the high-dimensional states have a higher transmission distance~\cite{cozzolino2019high}.
 This higher noise resilience of qudits is more advantageous if the qudits are entangled. The
entanglement becomes  more robust by increasing the dimension of the qudits while fixing their numbers. In other words,  as the noise sources act locally on every system, increasing the dimension $d$ will reduce the number of systems and thus reduce the effect of noise resulting in the robustness increase~\cite{liu2009decay}. The increasing noise level tolerance as the qudit dimension increases can be shown on an photonic OAM system as an example of its implementation~\cite{ecker2019overcoming}.

In summary the qudit system possesses advantages in the circuit design, physical implementation and has the potential to outperform the qubit system in various applications. 

\subsection{Future outlook of qudit system}
This review article introduces the basics of the high-dimensional qudit systems and provides details about qudit gates, qudit algorithms and implementations on various physical systems. 
The article serves as a summary of recent developments of qudit quantum computing and an introduction for newcomers to the field of qudit quantum computing.
Furthermore we show the advantages and the potential for qudit systems to outperform qubit counterparts. 
Of course these advantages can come with challenges such as possibly harder-to-implement universal gates, benchmarking~\cite{JWSS20,morvan2020qutrit,kononenko2020characterization}, characterization of qudit gate~\cite{reich2014optimal,gualdi2014efficient} and error correction connected with the complexity of the Clifford hierarchy for qudits~\cite{Web16}.

Compared to qubit systems, qudit systems currently have received less attention in both theoretical and experimental studies.
However, qudit quantum computing is becoming increasingly important as many 
topics and problems in this field are ripe for exploration.
Extending from qubits to qudits ushes in some mathematical challenges, with these mathematical problems elegant and perhaps giving new insights into quantum computing in their own right.
Connections between quantum resources such as entanglement, quantum algorithms and their improvements, scaling up qudit systems both to higher dimension and to more particles, benchmarking and error correction, and the bridging between qudits and continuous-variable quantum computing~\cite{GKP01} are examples of the fantastic research directions in this field of high-dimensional quantum computing.

\section*{Author Contributions}
All authors discussed the relevant materials to be added and all participated in writing the article.
\section*{Funding}
We would like to acknowledge the financial support  by the National Science Foundation under award number 1839191-ECCS.
BCS appreciates financial support from NSERC and from the Alberta Government.

\newpage
\bibliographystyle{ieeetr}
\bibliography{References}

\begin{thebibliography}{100}

\bibitem{brylinski2002universal}
J.-L. Brylinski and R.~Brylinski, ``Universal quantum gates,'' in {\em
  Mathematics of quantum computation}, pp.~117--134, Chapman and Hall/CRC,
  2002.

\bibitem{Hsuan-Hao2019}
H.-H. Lu, Z.~Hu, M.~S. Alshaykh, A.~J. Moore, Y.~Wang, P.~Imany, A.~M. Weiner,
  and S.~Kais, ``Quantum phase estimation with time-frequency qudits in a
  single photon,'' {\em Adv. Quantum Technol.}, vol.~0, no.~0, p.~1900074,
  2019.

\bibitem{Luo2014}
M.~Luo and X.~Wang, ``Universal quantum computation with qudits,'' {\em Science
  China Physics, Mechanics {\&} Astronomy}, vol.~57, pp.~1712--1717, Sep 2014.

\bibitem{li2013qutrits}
B.~Li, Z.-H. Yu, and S.-M. Fei, ``Geometry of quantum computation with
  qutrits,'' {\em Sci. Rep.}, vol.~3, p.~2594, 2013.

\bibitem{luo2014qudits}
M.-X. Luo, X.-B. Chen, Y.-X. Yang, and X.~Wang, ``Geometry of quantum
  computation with qudits,'' {\em Sci. Rep.}, vol.~4, p.~4044, 2014.

\bibitem{ZE12}
V.~E. Zobov and A.~S. Ermilov, ``Implementation of a quantum adiabatic
  algorithm for factorization on two qudits,'' {\em J. Exp. Theor. Phys.},
  vol.~114, pp.~923--932, 07 2012.

\bibitem{ADS13}
M.~H. Amin, N.~G. Dickson, and P.~Smith, ``Adiabatic quantum optimization with
  qudits,'' {\em Quantum Inf. Process.}, vol.~12, pp.~1819--1829, 04 2013.

\bibitem{cui2015universal}
S.~X. Cui and Z.~Wang, ``Universal quantum computation with metaplectic
  anyons,'' {\em J. Math. Phys.}, vol.~56, no.~3, p.~032202, 2015.

\bibitem{cui2015universal2}
S.~X. Cui, S.-M. Hong, and Z.~Wang, ``Universal quantum computation with weakly
  integral anyons,'' {\em Quantum Information Processing}, vol.~14, no.~8,
  pp.~2687--2727, 2015.

\bibitem{Bocharov2016ternary}
A.~Bocharov, S.~X. Cui, M.~Roetteler, and K.~M. Svore, ``Improved quantum
  ternary arithmetic,'' {\em Quantum Info. Comput.}, vol.~16, p.~862–884,
  July 2016.

\bibitem{GEZK19}
X.~Gao, M.~Erhard, A.~Zeilinger, and M.~Krenn, ``Computer-inspired concept for
  high-dimensional multipartite quantum gates,'' {\em Phys. Rev. Lett.},
  vol.~125, p.~050501, Jul 2020.

\bibitem{BdGS02}
S.~D. Bartlett, H.~de~Guise, and B.~C. Sanders, ``Quantum encodings in spin
  systems and harmonic oscillators,'' {\em Phys. Rev. A}, vol.~65, p.~052316,
  May 2002.

\bibitem{AHS16}
M.~R.~A. Adcock, P.~H{\o}yer, and B.~C. Sanders, ``Quantum computation with
  coherent spin states and the close {H}adamard problem,'' {\em Quantum Inf.
  Process.}, vol.~15, pp.~1361--1386, Jan 2016.

\bibitem{klimov2003}
A.~B. Klimov, R.~Guzm\'an, J.~C. Retamal, and C.~Saavedra, ``Qutrit quantum
  computer with trapped ions,'' {\em Phys. Rev. A}, vol.~67, p.~062313, Jun
  2003.

\bibitem{DOGRA2014}
S.~Dogra, Arvind, and K.~Dorai, ``Determining the parity of a permutation using
  an experimental nmr qutrit,'' {\em Phys. Lett. A}, vol.~378, no.~46, pp.~3452
  -- 3456, 2014.

\bibitem{Gedik2015}
Z.~Gedik, I.~A. Silva, B.~Lie~akmak, G.~Karpat, E.~L.~G. Vidoto, D.~O.
  Soares-Pinto, E.~R. deAzevedo, and F.~F. Fanchini, ``Computational speed-up
  with a single qudit,'' {\em Sci. Rep.}, vol.~5, pp.~14671 EP --, Oct 2015.
\newblock Article.

\bibitem{leuenberger2001}
M.~N. Leuenberger and D.~Loss, ``Quantum computing in molecular magnets,'' {\em
  Nature}, vol.~410, no.~6830, p.~789, 2001.

\bibitem{Howard2012}
M.~Howard and J.~Vala, ``Qudit versions of the qubit $\ensuremath{\pi}/8$
  gate,'' {\em Phys. Rev. A}, vol.~86, p.~022316, Aug 2012.

\bibitem{DWS03}
J.~Daboul, X.~Wang, and B.~C. Sanders, ``Quantum gates on hybrid qudits,'' {\em
  J. Phys. A.}, vol.~36, pp.~2525--2536, feb 2003.

\bibitem{Garcia-Escartin2013}
J.~C. Garcia-Escartin and P.~Chamorro-Posada, ``A swap gate for qudits,'' {\em
  Quantum Inf. Process.}, vol.~12, pp.~3625--3631, Dec 2013.

\bibitem{Ralph2007}
T.~C. Ralph, K.~J. Resch, and A.~Gilchrist, ``Efficient {Toffoli} gates using
  qudits,'' {\em Phys. Rev. A}, vol.~75, p.~022313, Feb 2007.

\bibitem{KNX+20}
E.~O. Kiktenko, A.~S. Nikolaeva, P.~Xu, G.~V. Shlyapnikov, and A.~K. Fedorov,
  ``Scalable quantum computing with qudits on a graph,'' {\em Phys. Rev. A},
  vol.~101, p.~022304, Feb 2020.

\bibitem{Nguyen2019}
D.~M. Nguyen and S.~Kim, ``Quantum key distribution protocol based on modified
  generalization of {Deutsch-Jozsa} algorithm in d-level quantum system,'' {\em
  Int. J. Theor. Phys}, vol.~58, pp.~71--82, Jan 2019.

\bibitem{NGP+20}
K.~Nagata, H.~Geurdes, S.~K. Patro, S.~Heidari, A.~Farouk, and T.~Nakamura,
  ``Generalization of the {Bernstein–Vazirani} algorithm beyond qubit
  systems,'' {\em Quantum Stud.: Math. Found.}, vol.~7, pp.~17--21, 03 2020.

\bibitem{Cao2011}
Y.~Cao, S.-G. Peng, C.~Zheng, and G.-L. Long, ``Quantum fourier transform and
  phase estimation in qudit system,'' {\em Commun. Theor. Phys}, vol.~55,
  pp.~790--794, may 2011.

\bibitem{BRS17}
A.~Bocharov, M.~Roetteler, and K.~M. Svore, ``Factoring with qutrits: Shor's
  algorithm on ternary and metaplectic quantum architectures,'' {\em Phys. Rev.
  A}, vol.~96, p.~012306, Jul 2017.

\bibitem{Ivanov2012}
S.~S. Ivanov, H.~S. Tonchev, and N.~V. Vitanov, ``Time-efficient implementation
  of quantum search with qudits,'' {\em Phys. Rev. A}, vol.~85, p.~062321, Jun
  2012.

\bibitem{vlasov2002}
A.~Y. Vlasov, ``Noncommutative tori and universal sets of nonbinary quantum
  gates,'' {\em J. Math. Phys.}, vol.~43, no.~6, pp.~2959--2964, 2002.

\bibitem{divincenzo1995two}
D.~P. DiVincenzo, ``Two-bit gates are universal for quantum computation,'' {\em
  Phys. Rev. A}, vol.~51, no.~2, p.~1015, 1995.

\bibitem{Got99}
D.~Gottesman, ``Fault-tolerant computation with higher-dimensional systems,''
  in {\em Lecture Notes in Computer Science} (W.~C.P., ed.), vol.~1509 of {\em
  NASA International Conference on Quantum Computing and Quantum
  Communications}, pp.~302--313, Berlin: Springer, 1999.

\bibitem{Zhou2003}
D.~L. Zhou, B.~Zeng, Z.~Xu, and C.~P. Sun, ``Quantum computation based on
  d-level cluster state,'' {\em Phys. Rev. A}, vol.~68, p.~062303, Dec 2003.

\bibitem{Bullock2005}
S.~S. Bullock, D.~P. O'Leary, and G.~K. Brennen, ``Asymptotically optimal
  quantum circuits for $d$-level systems,'' {\em Phys. Rev. Lett.}, vol.~94,
  p.~230502, Jun 2005.

\bibitem{Wen-Dong2013}
W.-D. Li, Y.-J. Gu, K.~Liu, Y.-H. Lee, and Y.-Z. Zhang, ``Efficient universal
  quantum computation with auxiliary {H}ilbert space,'' {\em Phys. Rev. A},
  vol.~88, p.~034303, Sep 2013.

\bibitem{Mischuck2013}
B.~Mischuck and K.~M\o{}lmer, ``Qudit quantum computation in the
  {J}aynes-{C}ummings model,'' {\em Phys. Rev. A}, vol.~87, p.~022341, Feb
  2013.

\bibitem{Reck1994}
M.~Reck, A.~Zeilinger, H.~J. Bernstein, and P.~Bertani, ``Experimental
  realization of any discrete unitary operator,'' {\em Phys. Rev. Lett.},
  vol.~73, pp.~58--61, Jul 1994.

\bibitem{RSdG99}
D.~J. Rowe, B.~C. Sanders, and H.~de~Guise, ``Representations of the {Weyl}
  group and {Wigner} functions for {SU(3)},'' {\em J. Math. Phys.}, vol.~40,
  no.~7, pp.~3604--3615, 1999.

\bibitem{Muthukrishnan2000}
A.~Muthukrishnan and C.~R. Stroud, ``Multivalued logic gates for quantum
  computation,'' {\em Phys. Rev. A}, vol.~62, p.~052309, Oct 2000.

\bibitem{Brennen2005}
G.~K. Brennen, D.~P. O'Leary, and S.~S. Bullock, ``Criteria for exact qudit
  universality,'' {\em Phys. Rev. A}, vol.~71, p.~052318, May 2005.

\bibitem{Got98}
D.~Gottesman, ``Theory of fault-tolerant quantum computation,'' {\em Phys. Rev.
  A}, vol.~57, pp.~127--137, Jan 1998.

\bibitem{BOYKIN2000}
P.~Boykin, T.~Mor, M.~Pulver, V.~Roychowdhury, and F.~Vatan, ``A new universal
  and fault-tolerant quantum basis,'' {\em Inf. Process. Lett.}, vol.~75,
  no.~3, pp.~101 -- 107, 2000.

\bibitem{PZ88}
J.~Patera and H.~Zassenhaus, ``The {Pauli} matrices in n dimensions and finest
  gradings of simple lie algebras of type {A} n- 1,'' {\em J. Math. Phys.},
  vol.~29, no.~3, pp.~665--673, 1988.

\bibitem{GKP01}
D.~Gottesman, A.~Kitaev, and J.~Preskill, ``Encoding a qubit in an
  oscillator,'' {\em Phys. Rev. A}, vol.~64, no.~1, p.~012310, 2001.

\bibitem{NBD+02}
M.~A. Nielsen, M.~J. Bremner, J.~L. Dodd, A.~M. Childs, and C.~M. Dawson,
  ``Universal simulation of hamiltonian dynamics for quantum systems with
  finite-dimensional state spaces,'' {\em Phys. Rev. A}, vol.~66, p.~022317,
  Aug 2002.

\bibitem{Web16}
Z.~Webb, ``The {C}lifford group forms a unitary 3-design,'' {\em Quantum Inf.
  Comput.}, vol.~16, no.~15{\&}16, pp.~1379--1400, 2016.

\bibitem{Gottesman1999}
D.~Gottesman and I.~L. Chuang, ``Demonstrating the viability of universal
  quantum computation using teleportation and single-qubit operations,'' {\em
  Nature}, vol.~402, no.~6760, pp.~390--393, 1999.

\bibitem{Eastin2009}
B.~Eastin and E.~Knill, ``Restrictions on transversal encoded quantum gate
  sets,'' {\em Phys. Rev. Lett.}, vol.~102, p.~110502, Mar 2009.

\bibitem{Zeng2007}
B.~Zeng, H.~Chung, A.~W. Cross, and I.~L. Chuang, ``Local unitary versus local
  {Clifford} equivalence of stabilizer and graph states,'' {\em Phys. Rev. A},
  vol.~75, p.~032325, Mar 2007.

\bibitem{Low2009}
R.~A. Low, ``Learning and testing algorithms for the {Clifford} group,'' {\em
  Phys. Rev. A}, vol.~80, p.~052314, Nov 2009.

\bibitem{Childs2001}
A.~M. Childs, ``Secure assisted quantum computation,'' {\em
  arXiv:quant-ph/0111046}, 2001.

\bibitem{vanDam2011}
W.~van Dam and M.~Howard, ``Noise thresholds for higher-dimensional systems
  using the discrete wigner function,'' {\em Phys. Rev. A}, vol.~83, p.~032310,
  Mar 2011.

\bibitem{ACB12}
H.~Anwar, E.~T. Campbell, and D.~E. Browne, ``Qutrit magic state
  distillation,'' {\em New J. Phys}, vol.~14, p.~063006, jun 2012.

\bibitem{CAB12}
E.~T. Campbell, H.~Anwar, and D.~E. Browne, ``Magic-state distillation in all
  prime dimensions using quantum {R}eed-{M}uller codes,'' {\em Phys. Rev. X},
  vol.~2, p.~041021, Dec 2012.

\bibitem{WILMOTT2011}
C.~M. Wilmott, ``On swapping the states of two qudits,'' {\em Int. J. Quantum
  Inf.}, vol.~09, no.~06, pp.~1511--1517, 2011.

\bibitem{WILMOTT2012}
C.~M. Wilmott and P.~R. Wild, ``On a generalized quantum swap gate,'' {\em Int.
  J. Quantum Inf.}, vol.~10, no.~03, p.~1250034, 2012.

\bibitem{Mermin2001}
N.~D. Mermin, ``From classical state swapping to quantum teleportation,'' {\em
  Phys. Rev. A}, vol.~65, p.~012320, Dec 2001.

\bibitem{Fujii_2003}
K.~Fujii, ``Exchange gate on the qudit space and fock space,'' {\em J. Opt. B},
  vol.~5, pp.~S613--S618, oct 2003.

\bibitem{Paz-Silva2009}
G.~A. Paz-Silva, S.~Rebi\ifmmode~\acute{c}\else \'{c}\fi{}, J.~Twamley, and
  T.~Duty, ``Perfect mirror transport protocol with higher dimensional quantum
  chains,'' {\em Phys. Rev. Lett.}, vol.~102, p.~020503, Jan 2009.

\bibitem{Wang_2001}
X.~Wang, ``Continuous-variable and hybrid quantum gates,'' {\em J. Phys. A},
  vol.~34, pp.~9577--9584, oct 2001.

\bibitem{Alber_2001}
G.~Alber, A.~Delgado, N.~Gisin, and I.~Jex, ``Efficient bipartite quantum state
  purification in arbitrary dimensional {H}ilbert spaces,'' {\em J. Phys. A},
  vol.~34, pp.~8821--8833, oct 2001.

\bibitem{Muthukrishnan2002}
A.~M. C.~R. Stroud, ``Quantum fast fourier transform using multilevel atoms,''
  {\em J. Mod. Opt.}, vol.~49, no.~13, pp.~2115--2127, 2002.

\bibitem{Cory1998}
D.~G. Cory, M.~D. Price, W.~Maas, E.~Knill, R.~Laflamme, W.~H. Zurek, T.~F.
  Havel, and S.~S. Somaroo, ``Experimental quantum error correction,'' {\em
  Phys. Rev. Lett.}, vol.~81, pp.~2152--2155, Sep 1998.

\bibitem{Dennis2001}
E.~Dennis, ``Toward fault-tolerant quantum computation without concatenation,''
  {\em Phys. Rev. A}, vol.~63, p.~052314, Apr 2001.

\bibitem{shi2002}
Y.~Shi, ``Both {Toffoli} and controlled-{NOT} need little help to do universal
  quantum computation,'' {\em arXiv preprint quant-ph/0205115}, 2002.

\bibitem{Nielsen2011}
M.~A. Nielsen and I.~L. Chuang, {\em Quantum Computation and Quantum
  Information: 10th Anniversary Edition}.
\newblock New York, NY, USA: Cambridge University Press, 10th~ed., 2011.

\bibitem{Aspuru-Guzik1704}
A.~Aspuru-Guzik, A.~D. Dutoi, P.~J. Love, and M.~Head-Gordon, ``Simulated
  quantum computation of molecular energies,'' {\em Science}, vol.~309,
  no.~5741, pp.~1704--1707, 2005.

\bibitem{shor1994}
P.~W. {Shor}, ``Algorithms for quantum computation: discrete logarithms and
  factoring,'' in {\em Proceedings 35th Annual Symposium on Foundations of
  Computer Science}, pp.~124--134, Nov 1994.

\bibitem{CEMM98}
R.~Cleve, A.~Ekert, C.~Macchiavello, and M.~Mosca, ``Quantum algorithms
  revisited,'' {\em Proceedings of the Royal Society of London. Series A:
  Mathematical, Physical and Engineering Sciences}, vol.~454, no.~1969,
  pp.~339--354, 1998.

\bibitem{Di2013}
Y.-M. Di and H.-R. Wei, ``Synthesis of multivalued quantum logic circuits by
  elementary gates,'' {\em Phys. Rev. A}, vol.~87, p.~012325, Jan 2013.

\bibitem{KHAN2006336}
F.~S. Khan and M.~Perkowski, ``Synthesis of multi-qudit hybrid and d-valued
  quantum logic circuits by decomposition,'' {\em Theor. Comput. Sci.},
  vol.~367, no.~3, pp.~336 -- 346, 2006.

\bibitem{nielsen2006quantum}
M.~A. Nielsen, M.~R. Dowling, M.~Gu, and A.~C. Doherty, ``Quantum computation
  as geometry,'' {\em Science}, vol.~311, no.~5764, pp.~1133--1135, 2006.

\bibitem{nielsen2005geometric}
M.~A. Nielsen, ``A geometric approach to quantum circuit lower bounds,'' {\em
  arXiv preprint quant-ph/0502070}, 2005.

\bibitem{Deu85}
D.~Deutsch, ``Quantum theory, the {C}hurch--{T}uring principle and the
  universal quantum computer,'' {\em J. Phys. A}, vol.~400, no.~1818,
  pp.~97--117, 1985.

\bibitem{DJ92}
D.~P. Deutsch and R.~Jozsa, ``Rapid solution of problems by quantum
  computation,'' {\em Proc. Royal Soc. Lond. A}, vol.~439, pp.~553--558, 1992.

\bibitem{Van_den_Nest2013}
M.~Van~den Nest, ``Universal quantum computation with little entanglement,''
  {\em Phys. Rev. Lett.}, vol.~110, p.~060504, Feb 2013.

\bibitem{Howard2014}
M.~Howard, J.~Wallman, V.~Veitch, and J.~Emerson, ``Contextuality supplies the
  'magic' for quantum computation,'' {\em Nature}, vol.~510, p.~351, Jun 2014.
\newblock Article.

\bibitem{Kochen1975}
S.~Kochen and E.~P. Specker, {\em The Problem of Hidden Variables in Quantum
  Mechanics}, pp.~293--328.
\newblock Dordrecht: Springer Netherlands, 1975.

\bibitem{Klyachko2008}
A.~A. Klyachko, M.~A. Can, S.~Binicio\ifmmode~\breve{g}\else \u{g}\fi{}lu, and
  A.~S. Shumovsky, ``Simple test for hidden variables in spin-1 systems,'' {\em
  Phys. Rev. Lett.}, vol.~101, p.~020403, Jul 2008.

\bibitem{Zhan2015}
X.~Zhan, J.~Li, H.~Qin, Z.~hao Bian, and P.~Xue, ``Linear optical demonstration
  of quantum speed-up with a single qudit,'' {\em Opt. Express}, vol.~23,
  pp.~18422--18427, Jul 2015.

\bibitem{Fan2007}
Y.~Fan, ``A generalization of the {D}eutsch-{J}ozsa algorithm to multi-valued
  quantum logic,'' in {\em 37th International Symposium on Multiple-Valued
  Logic (ISMVL'07)}, pp.~12--12, May 2007.

\bibitem{KIKTENKO2015}
E.~Kiktenko, A.~Fedorov, A.~Strakhov, and V.~Man'ko, ``Single qudit realization
  of the {D}eutsch algorithm using superconducting many-level quantum
  circuits,'' {\em Phys. Lett. A}, vol.~379, no.~22, pp.~1409 -- 1413, 2015.

\bibitem{Kessel2002}
A.~R. Kessel and N.~M. Yakovleva, ``Implementation schemes in {NMR} of quantum
  processors and the {D}eutsch-{J}ozsa algorithm by using virtual spin
  representation,'' {\em Phys. Rev. A}, vol.~66, p.~062322, Dec 2002.

\bibitem{Bernstein1997}
E.~Bernstein and U.~Vazirani, ``Quantum complexity theory,'' {\em SIAM Journal
  on Computing}, vol.~26, no.~5, pp.~1411--1473, 1997.

\bibitem{KMS16}
R.~Krishna, V.~Makwana, and A.~P. Suresh, ``A generalization of
  {B}ernstein-{V}azirani algorithm to qudit systems,'' {\em arXiv preprint
  arXiv:1609.03185}, 2016.

\bibitem{Zilic2007}
Z.~{Zilic} and K.~{Radecka}, ``Scaling and better approximating quantum
  {Fourier} transform by higher radices,'' {\em IEEE Transactions on
  Computers}, vol.~56, pp.~202--207, Feb 2007.

\bibitem{Coppersmith1994}
D.~Coppersmith, ``An approximate {F}ourier transform useful in quantum
  factoring,'' {\em arXiv preprint quant-ph/0201067}, 2002.

\bibitem{Shende2006}
V.~V. {Shende}, S.~S. {Bullock}, and I.~L. {Markov}, ``Synthesis of
  quantum-logic circuits,'' {\em IEEE Transactions on Computer-Aided Design of
  Integrated Circuits and Systems}, vol.~25, pp.~1000--1010, June 2006.

\bibitem{Parasa2011}
V.~{Parasa} and M.~{Perkowski}, ``Quantum phase estimation using multivalued
  logic,'' in {\em 2011 41st IEEE International Symposium on Multiple-Valued
  Logic}, pp.~224--229, May 2011.

\bibitem{Lloyd1999}
D.~S. Abrams and S.~Lloyd, ``Quantum algorithm providing exponential speed
  increase for finding eigenvalues and eigenvectors,'' {\em Phys. Rev. Lett.},
  vol.~83, pp.~5162--5165, Dec 1999.

\bibitem{Lloyd2009}
A.~W. Harrow, A.~Hassidim, and S.~Lloyd, ``Quantum algorithm for linear systems
  of equations,'' {\em Phys. Rev. Lett.}, vol.~103, p.~150502, Oct 2009.

\bibitem{pan2014experimental}
J.~Pan, Y.~Cao, X.~Yao, Z.~Li, C.~Ju, H.~Chen, X.~Peng, S.~Kais, and J.~Du,
  ``Experimental realization of quantum algorithm for solving linear systems of
  equations,'' {\em Phys. Rev. A}, vol.~89, no.~2, p.~022313, 2014.

\bibitem{Tonchev2016}
H.~S. Tonchev and N.~V. Vitanov, ``Quantum phase estimation and quantum
  counting with qudits,'' {\em Phys. Rev. A}, vol.~94, p.~042307, Oct 2016.

\bibitem{Hefeng2008}
H.~Wang, S.~Kais, A.~Aspuru-Guzik, and M.~R. Hoffmann, ``Quantum algorithm for
  obtaining the energy spectrum of molecular systems,'' {\em Phys. Chem. Chem.
  Phys.}, vol.~10, pp.~5388--5393, 2008.

\bibitem{daskin2011decomposition}
A.~Daskin and S.~Kais, ``Decomposition of unitary matrices for finding quantum
  circuits: application to molecular hamiltonians,'' {\em J. Chem. Phys},
  vol.~134, no.~14, p.~144112, 2011.

\bibitem{daskin2012universal}
A.~Daskin, A.~Grama, G.~Kollias, and S.~Kais, ``Universal programmable quantum
  circuit schemes to emulate an operator,'' {\em J. Chem. Phys}, vol.~137,
  no.~23, p.~234112, 2012.

\bibitem{kais2014introduction}
S.~Kais, ``Introduction to quantum information and computation for chemistry,''
  {\em Quantum Inf. Comput. Chem}, vol.~154, pp.~1--38, 2014.

\bibitem{bian2019quantum}
T.~Bian, D.~Murphy, R.~Xia, A.~Daskin, and S.~Kais, ``Quantum computing methods
  for electronic states of the water molecule,'' {\em Mol. Phys}, vol.~117,
  no.~15-16, pp.~2069--2082, 2019.

\bibitem{sawerwain2013quantum}
M.~Sawerwain and W.~Leo{\'n}ski, ``Quantum circuits based on qutrits as a tool
  for solving systems of linear equations,'' {\em arXiv:1309.0800}, 2013.

\bibitem{shor1999polynomial}
P.~W. Shor, ``Polynomial-time algorithms for prime factorization and discrete
  logarithms on a quantum computer,'' {\em SIAM review}, vol.~41, no.~2,
  pp.~303--332, 1999.

\bibitem{fan2008}
Y.~Fan, ``Applications of multi-valued quantum algorithms,'' {\em
  arXiv:0809.0932}, 2008.

\bibitem{LI20114249}
H.~Li, C.~Wu, W.~Liu, P.~Chen, and C.~Li, ``Fast quantum search algorithm for
  databases of arbitrary size and its implementation in a cavity {QED}
  system,'' {\em Phys. Lett. A}, vol.~375, no.~48, pp.~4249 -- 4254, 2011.

\bibitem{Kyoseva2006}
E.~S. Kyoseva and N.~V. Vitanov, ``Coherent pulsed excitation of degenerate
  multistate systems: Exact analytic solutions,'' {\em Phys. Rev. A}, vol.~73,
  p.~023420, Feb 2006.

\bibitem{RB01}
R.~Raussendorf and H.~J. Briegel, ``A one-way quantum computer,'' {\em Phys.
  Rev. Lett.}, vol.~86, pp.~5188--5191, May 2001.

\bibitem{FGGS00}
E.~Farhi, J.~Goldstone, S.~Gutmann, and M.~Sipser, ``Quantum computation by
  adiabatic evolution,'' {\em arXiv:quant-ph/0001106}, 2000.

\bibitem{AvDK+07}
D.~Aharonov, W.~van Dam, J.~Kempe, Z.~Landau, S.~Lloyd, and O.~Regev,
  ``Adiabatic quantum computation is equivalent to standard quantum
  computation,'' {\em SIAM J. Comput.}, vol.~37, no.~1, pp.~166--194, 2007.

\bibitem{FLW02}
M.~H. Freedman, M.~Larsen, and Z.~Wang, ``A modular functor which is
  universal¶for quantum computation,'' {\em Commun. Math. Phys.}, vol.~227,
  pp.~605--622, 06 2002.

\bibitem{BR01}
H.~J. Briegel and R.~Raussendorf, ``Persistent entanglement in arrays of
  interacting particles,'' {\em Phys. Rev. Lett.}, vol.~86, pp.~910--913, Jan
  2001.

\bibitem{HEB04}
M.~Hein, J.~Eisert, and H.~J. Briegel, ``Multiparty entanglement in graph
  states,'' {\em Phys. Rev. A}, vol.~69, p.~062311, Jun 2004.

\bibitem{Nie06}
M.~A. Nielsen, ``Cluster-state quantum computation,'' {\em Rep. Math. Phys.},
  vol.~57, pp.~147--161, 02 2006.

\bibitem{KFMS10}
A.~Keet, B.~Fortescue, D.~Markham, and B.~C. Sanders, ``Quantum secret sharing
  with qudit graph states,'' {\em Phys. Rev. A}, vol.~82, p.~062315, Dec 2010.

\bibitem{JLKK19}
J.~Joo, C.-W. Lee, S.~Kono, and J.~Kim, ``Logical measurement-based quantum
  computation in circuit-qed,'' {\em Sci. Rep.}, vol.~9, p.~16592, 11 2019.

\bibitem{FGS+94}
A.~B. Finnila, M.~A. Gomez, C.~Sebenik, C.~Stenson, and J.~D. Doll, ``Quantum
  annealing: A new method for minimizing multidimensional functions,'' {\em
  Chem. Phys. Lett.}, vol.~219, no.~5, pp.~343 -- 348, 1994.

\bibitem{KN98}
T.~Kadowaki and H.~Nishimori, ``Quantum annealing in the transverse ising
  model,'' {\em Phys. Rev. E}, vol.~58, pp.~5355--5363, Nov 1998.

\bibitem{DC08}
A.~Das and B.~K. Chakrabarti, ``Colloquium: Quantum annealing and analog
  quantum computation,'' {\em Rev. Mod. Phys.}, vol.~80, pp.~1061--1081, Sep
  2008.

\bibitem{BKM+14}
R.~Barends, J.~Kelly, A.~Megrant, A.~Veitia, D.~Sank, E.~Jeffrey, T.~C. White,
  J.~Mutus, A.~G. Fowler, B.~Campbell, Y.~Chen, Z.~Chen, B.~Chiaro,
  A.~Dunsworth, C.~Neill, P.~O’Malley, P.~Roushan, A.~Vainsencher, J.~Wenner,
  A.~N. Korotkov, A.~N. Cleland, and J.~M. Martinis, ``Superconducting quantum
  circuits at the surface code threshold for fault tolerance,'' {\em Nature},
  vol.~508, pp.~500--503, 04 2014.

\bibitem{MG14}
J.~M. Martinis and M.~R. Geller, ``Fast adiabatic qubit gates using only
  ${\ensuremath{\sigma}}_{z}$ control,'' {\em Phys. Rev. A}, vol.~90,
  p.~022307, Aug 2014.

\bibitem{ZGS15}
E.~Zahedinejad, J.~Ghosh, and B.~C. Sanders, ``High-fidelity single-shot
  {T}offoli gate via quantum control,'' {\em Phys. Rev. Lett.}, vol.~114,
  p.~200502, May 2015.

\bibitem{ZGS16}
E.~Zahedinejad, J.~Ghosh, and B.~C. Sanders, ``Designing high-fidelity
  single-shot three-qubit gates: A machine-learning approach,'' {\em Phys. Rev.
  Applied}, vol.~6, p.~054005, Nov 2016.

\bibitem{WSK20}
S.~Watabe, Y.~Seki, and S.~Kawabata, ``Enhancing quantum annealing performance
  by a degenerate two-level system,'' {\em Sci. Rep.}, vol.~10, p.~146, 12
  2020.

\bibitem{HL16}
A.~Hutter and D.~Loss, ``Quantum computing with parafermions,'' {\em Phys. Rev.
  B}, vol.~93, p.~125105, Mar 2016.

\bibitem{DMCJ19}
A.~Dua, B.~Malomed, M.~Cheng, and L.~Jiang, ``Universal quantum computing with
  parafermions assisted by a half-fluxon,'' {\em Phys. Rev. B}, vol.~100,
  p.~144508, Oct 2019.

\bibitem{Kit03}
A.~Kitaev, ``Fault-tolerant quantum computation by anyons,'' {\em Ann. Phys.
  (N. Y.)}, vol.~303, no.~1, pp.~2 -- 30, 2003.

\bibitem{Jelezko2004}
F.~Jelezko, T.~Gaebel, I.~Popa, M.~Domhan, A.~Gruber, and J.~Wrachtrup,
  ``Observation of coherent oscillation of a single nuclear spin and
  realization of a two-qubit conditional quantum gate,'' {\em Phys. Rev.
  Lett.}, vol.~93, p.~130501, Sep 2004.

\bibitem{Childress2006}
L.~Childress, M.~V. Gurudev~Dutt, J.~M. Taylor, A.~S. Zibrov, F.~Jelezko,
  J.~Wrachtrup, P.~R. Hemmer, and M.~D. Lukin, ``Coherent dynamics of coupled
  electron and nuclear spin qubits in diamond,'' {\em Science}, vol.~314,
  no.~5797, pp.~281--285, 2006.

\bibitem{Neumann2010}
P.~Neumann, J.~Beck, M.~Steiner, F.~Rempp, H.~Fedder, P.~R. Hemmer,
  J.~Wrachtrup, and F.~Jelezko, ``Single-shot readout of a single nuclear
  spin,'' {\em Science}, vol.~329, no.~5991, pp.~542--544, 2010.

\bibitem{Loss1998}
D.~Loss and D.~P. DiVincenzo, ``Quantum computation with quantum dots,'' {\em
  Phys. Rev. A}, vol.~57, pp.~120--126, Jan 1998.

\bibitem{Nowack2007}
K.~C. Nowack, F.~H.~L. Koppens, Y.~V. Nazarov, and L.~M.~K. Vandersypen,
  ``Coherent control of a single electron spin with electric fields,'' {\em
  Science}, vol.~318, no.~5855, pp.~1430--1433, 2007.

\bibitem{Milburn_2009}
G.~J. Milburn, ``Photons as qubits,'' {\em Physica Scripta}, vol.~T137,
  p.~014003, dec 2009.

\bibitem{Prevedel2007}
R.~Prevedel, P.~Walther, F.~Tiefenbacher, P.~BMei~hi, R.~Kaltenbaek,
  T.~Jennewein, and A.~Zeilinger, ``High-speed linear optics quantum computing
  using active feed-forward,'' {\em Nature}, vol.~445, no.~7123, pp.~65--69,
  2007.

\bibitem{Chiorescu2003}
I.~Chiorescu, Y.~Nakamura, C.~J. P.~M. Harmans, and J.~E. Mooij, ``Coherent
  quantum dynamics of a superconducting flux qubit,'' {\em Science}, vol.~299,
  no.~5614, pp.~1869--1871, 2003.

\bibitem{Clarke2008}
J.~Clarke and F.~K. Wilhelm, ``Superconducting quantum bits,'' {\em Nature},
  vol.~453, no.~7198, pp.~1031--1042, 2008.

\bibitem{Blatt2008}
R.~Blatt and D.~Wineland, ``Entangled states of trapped atomic ions,'' {\em
  Nature}, vol.~453, no.~7198, pp.~1008--1015, 2008.

\bibitem{Bloch2008}
I.~Bloch, ``Quantum coherence and entanglement with ultracold atoms in optical
  lattices,'' {\em Nature}, vol.~453, no.~7198, pp.~1016--1022, 2008.

\bibitem{troiani2011}
F.~Troiani and M.~Affronte, ``Molecular spins for quantum information
  technologies,'' {\em Chem. Soc. Rev}, vol.~40, no.~6, pp.~3119--3129, 2011.

\bibitem{bogani2010}
L.~Bogani and W.~Wernsdorfer, ``Molecular spintronics using single-molecule
  magnets,'' in {\em Nanoscience And Technology: A Collection of Reviews from
  Nature Journals}, pp.~194--201, World Scientific, 2010.

\bibitem{clemente2012magnetic}
J.~M. Clemente-Juan, E.~Coronado, and A.~Gaita-Ari{\~n}o, ``Magnetic
  polyoxometalates: from molecular magnetism to molecular spintronics and
  quantum computing,'' {\em Chem. Soc. Rev}, vol.~41, no.~22, pp.~7464--7478,
  2012.

\bibitem{aromi2012design}
G.~Arom{\'\i}, D.~Aguila, P.~Gamez, F.~Luis, and O.~Roubeau, ``Design of
  magnetic coordination complexes for quantum computing,'' {\em Chem. Soc.
  Rev}, vol.~41, no.~2, pp.~537--546, 2012.

\bibitem{chuang1998experimental}
I.~L. Chuang, L.~M. Vandersypen, X.~Zhou, D.~W. Leung, and S.~Lloyd,
  ``Experimental realization of a quantum algorithm,'' {\em Nature}, vol.~393,
  no.~6681, p.~143, 1998.

\bibitem{vandersypen2001experimental}
L.~M. Vandersypen, M.~Steffen, G.~Breyta, C.~S. Yannoni, M.~H. Sherwood, and
  I.~L. Chuang, ``Experimental realization of {S}hor's quantum factoring
  algorithm using nuclear magnetic resonance,'' {\em Nature}, vol.~414,
  no.~6866, p.~883, 2001.

\bibitem{Moreno-Pineda2018}
E.~Moreno-Pineda, C.~Godfrin, F.~Balestro, W.~Wernsdorfer, and M.~Ruben,
  ``Molecular spin qudits for quantum algorithms,'' {\em Chem. Soc. Rev.},
  vol.~47, pp.~501--513, 2018.

\bibitem{Babazadeh2017}
A.~Babazadeh, M.~Erhard, F.~Wang, M.~Malik, R.~Nouroozi, M.~Krenn, and
  A.~Zeilinger, ``High-dimensional single-photon quantum gates: Concepts and
  experiments,'' {\em Phys. Rev. Lett.}, vol.~119, p.~180510, Nov 2017.

\bibitem{EFKZ18}
M.~Erhard, R.~Fickler, M.~Krenn, and A.~Zeilinger, ``Twisted photons: new
  quantum perspectives in high dimensions,'' {\em Light Sci. Appl}, vol.~7,
  p.~17146, 2018.

\bibitem{KRR+17}
M.~Kues, C.~Reimer, P.~Roztocki, L.~Romero~Cortés, S.~Sciara, B.~Wetzel,
  Y.~Zhang, A.~Cino, S.~T. Chu, B.~E. Little, D.~J. Moss, L.~Caspani,
  J.~Aza\~{n}a, and R.~Morandotti, ``On-chip generation of high-dimensional
  entangled quantum states and their coherent control,'' {\em Nature},
  vol.~546, no.~6760, pp.~622--626, 2017.

\bibitem{lu2018electro}
H.-H. Lu, J.~M. Lukens, N.~A. Peters, O.~D. Odele, D.~E. Leaird, A.~M. Weiner,
  and P.~Lougovski, ``Electro-optic frequency beam splitters and tritters for
  high-fidelity photonic quantum inf. process.,'' {\em Phys. Rev. Lett.},
  vol.~120, no.~3, p.~030502, 2018.

\bibitem{IJO+18}
P.~Imany, J.~A. Jaramillo-Villegas, O.~D. Odele, K.~Han, D.~E. Leaird, J.~M.
  Lukens, P.~Lougovski, M.~Qi, and A.~M. Weiner, ``50-ghz-spaced comb of
  high-dimensional frequency-bin entangled photons from an on-chip silicon
  nitride microresonator,'' {\em Opt. Express}, vol.~26, no.~2, pp.~1825--1840,
  2018.

\bibitem{IJA+19}
P.~Imany, J.~A. Jaramillo-Villegas, M.~S. Alshaykh, J.~M. Lukens, O.~D. Odele,
  A.~J. Moore, D.~E. Leaird, M.~Qi, and A.~M. Weiner, ``High-dimensional
  optical quantum logic in large operational spaces,'' {\em npj Quantum Inf.},
  vol.~5, Jul 2019.

\bibitem{islam2017}
N.~T. Islam, C.~C.~W. Lim, C.~Cahall, J.~Kim, and D.~J. Gauthier, ``Provably
  secure and high-rate quantum key distribution with time-bin qudits,'' {\em
  Sci. Adv}, vol.~3, no.~11, p.~e1701491, 2017.

\bibitem{Humphreys2013}
P.~C. Humphreys, B.~J. Metcalf, J.~B. Spring, M.~Moore, X.-M. Jin, M.~Barbieri,
  W.~S. Kolthammer, and I.~A. Walmsley, ``Linear optical quantum computing in a
  single spatial mode,'' {\em Phys. Rev. Lett.}, vol.~111, p.~150501, Oct 2013.

\bibitem{mc2005trapped}
D.~Mc~Hugh and J.~Twamley, ``Trapped-ion qutrit spin molecule quantum
  computer,'' {\em New J. Phys.}, vol.~7, no.~1, p.~174, 2005.

\bibitem{bovey1988nuclear}
F.~A. Bovey, P.~A. Mirau, and H.~Gutowsky, {\em Nuclear magnetic resonance
  spectroscopy}.
\newblock Elsevier, 1988.

\bibitem{slichter2013principles}
C.~P. Slichter, {\em Principles of magnetic resonance}, vol.~1.
\newblock Springer Science \& Business Media, 2013.

\bibitem{lee2006projective}
J.-S. Lee and A.~K. Khitrin, ``Projective measurement in nuclear magnetic
  resonance,'' {\em Appl. Phys. Lett}, vol.~89, no.~7, p.~074105, 2006.

\bibitem{Moreno2016}
E.~Moreno~Pineda, T.~Komeda, K.~Katoh, M.~Yamashita, and M.~Ruben, ``Surface
  confinement of {TbPc2-SMMs}: structural{,} electronic and magnetic
  properties,'' {\em Dalton Trans.}, vol.~45, pp.~18417--18433, 2016.

\bibitem{ganzhorn2013strong}
M.~Ganzhorn, S.~Klyatskaya, M.~Ruben, and W.~Wernsdorfer, ``Strong spin--phonon
  coupling between a single-molecule magnet and a carbon nanotube
  nanoelectromechanical system,'' {\em Nat. Nanotechnol.}, vol.~8, no.~3,
  p.~165, 2013.

\bibitem{urdampilleta2011}
M.~Urdampilleta, S.~Klyatskaya, J.-P. Cleuziou, M.~Ruben, and W.~Wernsdorfer,
  ``Supramolecular spin valves,'' {\em Nat. Mater}, vol.~10, no.~7, p.~502,
  2011.

\bibitem{thiele2014}
S.~Thiele, F.~Balestro, R.~Ballou, S.~Klyatskaya, M.~Ruben, and W.~Wernsdorfer,
  ``Electrically driven nuclear spin resonance in single-molecule magnets,''
  {\em Science}, vol.~344, no.~6188, pp.~1135--1138, 2014.

\bibitem{ishikawa2003}
N.~Ishikawa, M.~Sugita, T.~Okubo, N.~Tanaka, T.~Iino, and Y.~Kaizu,
  ``Determination of {l}igand-field parameters and f-electronic structures of
  double-decker bis (phthalocyaninato) lanthanide complexes,'' {\em Inorganic
  chemistry}, vol.~42, no.~7, pp.~2440--2446, 2003.

\bibitem{godfrin2017}
C.~Godfrin, A.~Ferhat, R.~Ballou, S.~Klyatskaya, M.~Ruben, W.~Wernsdorfer, and
  F.~Balestro, ``Operating quantum states in single magnetic molecules:
  implementation of grover’s quantum algorithm,'' {\em Phys. Rev. Lett.},
  vol.~119, no.~18, p.~187702, 2017.

\bibitem{cozzolino2019high}
D.~Cozzolino, B.~Da~Lio, D.~Bacco, and L.~K. Oxenl{\o}we, ``High-dimensional
  quantum communication: Benefits, progress, and future challenges,'' {\em Adv.
  Quantum Technol.}, vol.~2, no.~12, p.~1900038, 2019.

\bibitem{sheridan2010security}
L.~Sheridan and V.~Scarani, ``Security proof for quantum key distribution using
  qudit systems,'' {\em Phys. Rev. A}, vol.~82, no.~3, p.~030301, 2010.

\bibitem{liu2009decay}
Z.~Liu and H.~Fan, ``Decay of multiqudit entanglement,'' {\em Phys. Rev. A},
  vol.~79, no.~6, p.~064305, 2009.

\bibitem{ecker2019overcoming}
S.~Ecker, F.~Bouchard, L.~Bulla, F.~Brandt, O.~Kohout, F.~Steinlechner,
  R.~Fickler, M.~Malik, Y.~Guryanova, R.~Ursin, {\em et~al.}, ``Overcoming
  noise in entanglement distribution,'' {\em Phys. Rev. X}, vol.~9, no.~4,
  p.~041042, 2019.

\bibitem{JWSS20}
M.~Jafarzadeh, Y.-D. Wu, Y.~R. Sanders, and B.~C. Sanders, ``Randomized
  benchmarking for qudit {C}lifford gates,'' {\em New J. Phys.}, vol.~22,
  p.~063014, 06 2020.

\bibitem{morvan2020qutrit}
A.~Morvan, V.~Ramasesh, M.~Blok, J.~Kreikebaum, K.~O'Brien, L.~Chen,
  B.~Mitchell, R.~Naik, D.~Santiago, and I.~Siddiqi, ``Qutrit randomized
  benchmarking,'' {\em arXiv preprint arXiv:2008.09134}, 2020.

\bibitem{kononenko2020characterization}
M.~Kononenko, M.~Yurtalan, J.~Shi, and A.~Lupascu, ``Characterization of
  control in a superconducting qutrit using randomized benchmarking,'' {\em
  arXiv preprint arXiv:2009.00599}, 2020.

\bibitem{reich2014optimal}
D.~M. Reich, G.~Gualdi, and C.~P. Koch, ``Optimal qudit operator bases for
  efficient characterization of quantum gates,'' {\em J. Phys. A-Math Theor},
  vol.~47, no.~38, p.~385305, 2014.

\bibitem{gualdi2014efficient}
G.~Gualdi, D.~Licht, D.~M. Reich, and C.~P. Koch, ``Efficient monte carlo
  characterization of quantum operations for qudits,'' {\em Phys. Rev. A},
  vol.~90, no.~3, p.~032317, 2014.

\end{thebibliography}
\end{document}